\documentclass[%
 reprint,
superscriptaddress,
showpacs,preprintnumbers,
 amsmath,amssymb,
 aps,
 prb,
]{revtex4-2}
\usepackage{graphicx}
\usepackage{xcolor}
\usepackage{dcolumn}
\usepackage{bm}

\begin{document}

\title{Magnetic resonance in quantum computing and in accurate measurements of the nuclear moments of atoms and molecules}
\author{Zhichen Liu}
\affiliation{Department of Physics, University of Central Florida, Orlando, FL 32816-2385, USA}
\author{Richard A. Klemm}
\email{richard.klemm@ucf.edu, corresponding author}
\affiliation{Department of Physics, University of Central Florida, Orlando, FL 32816-2385, USA}
\affiliation{U. S. Air Force Research Laboratory, Wright-Patterson Air Force Base, Ohio 45433-7251, USA}
\date{\today}

\begin{abstract}
With modern experimental techniques, it is now possible to generate magnetic fields with the precise form, ${\bm H}(t)=H_0\hat{\bm z}+H_1[\hat{\bm x}\cos(\omega t)+\hat{\bm y}\sin(\omega t)]$,  both at frequencies within an order of magnitude of the nuclear magnetic resonance (NMR) and of the electron paramagnetic resonance (EPR) frequencies, $\omega_{n0}$ and $\omega_{e0}$, respectively, when acting on atoms or molecules.
We found  simple forms for the exact wave functions of the nuclear and electronic spins that allow for transitions between entangled  states, allowing an atom or molecule to function as a quantum computer.   These forms also allow for precise NMR or EPR measurements of the nuclear moments in atoms or molecules. Examples relevant to measurements of the nuclear moments of $^{14}$N, $^7$Li, and $^{133}$Cs are given.  Since present hyperfine measurements of the lowest three nuclear moments of $^{133}$Cs are inconsistent with each other, the proposed NMR or EPR experiments allow one to measure all seven of its nuclear moments with high precision.

\end{abstract}

 \vskip0pt
\maketitle

\section{Introduction}\
Nuclear magnetic  moments have spin quantum numbers $I$ that  for light stable  isotopes are $\frac{1}{2}$ for $^1$H and $^{13}$C, 1 for $^2$H and $^{14}$N, $\frac{3}{2}$ for $^7$Li, $^9$Be, and $^{11}$B, $\frac{5}{2}$ for $^{17}$O, 3 for $^{10}$B, and for heavy stable isotopes, it is $\frac{7}{2}$ for $^{133}$Cs and up to  $\frac{9}{2}$ for  $^{83}$Kr, $^{87}$Sr, $^{93}$Nb, $^{113}$In, $^{179}$Hf, and $^{209}$Bi, and to yet higher values for long-lived unstable nuclei \cite{Stone,CIAAW}.   As first shown by Ramsey \cite{Ramsey}, nuclei with spin $I$ can have $2I$ moments, with the matrix set $\hat{\delta V}_{ne}$ of those  nuclear  ${\bm I}$ and electronic  ${\bm J}$  spin angular momentum operators given by
\begin{eqnarray}
\hat{\delta V}_{ne} &=&\sum_{q=0}^{2I}K_q({\bm I}\cdot{\bm J}/\hbar^2)^{q}{\bm 1}_n\otimes{\bm 1}_e,\label{Delta}
\end{eqnarray}
where ${\bm 1}_n$ and ${\bm 1}_e$ are identity matrices of respective ranks $2I+1$ and $2J+1$,  the $K_q$ are real functions of $I$ and $J$ and the  components of ${\bm I}$ and of  ${\bm J}$ satisfy independent Lie algebras,  each of the form
\begin{eqnarray}
[O_{\lambda},O_{\mu}]&=&i\hbar\epsilon_{\lambda\mu\nu}O_{\nu},\label{commutator}
\end{eqnarray}
where $[I_{\lambda},J_{\mu}]=0$, $\hbar=\frac{h}{2\pi}$, $h$ is Planck's constant, $\epsilon_{\lambda\mu\nu}$ is the Levi-Civita symbol,  and repeated subscripts are implicitly summed over.
 Eq. (\ref{Delta}) implies that  nuclei with $I=\frac{1}{2}$ exibit magnetic dipole moments, those with $I=1$ have both magnetic dipole and electric quadrupole moments, and those with $I=\frac{3}{2}$ also  have magnetic octupole moments, etc. Explicit formulas for the $K_q$ for $0\le q\le3$ valid for those three moments are available \cite{Trees,Armstrong}.  However, for $I>\frac{3}{2}$, hexadecapole and yet higher moments can exist, and if so,  they will  contribute to the   $K_q$ for $q>3$ and also to the  $K_q$ expressions for $0\le q\le3$.  An important example discussed in the following is $^{133}$Cs, for which present hyperfine measurements of the lowest three of its seven nuclear moments have been at least somewhat inconsistent with one another  \cite{Allegrini}, possibly because all higher moments were assumed to vanish. 

Traditionally, such moments have been  easiest to measure by careful examinations of the electronic hyperfine structure of an atom \cite{Trees,Armstrong,AHSI,AHSII,AHSIII}.   In the following, we show that it is possible to make much more accurate measurements of the nuclear moments using either nuclear magnetic resonance (NMR) or electron paramagnetic resonance (EPR) experiments.   That is, by appropriate designs of solenoids,  magnetic fields can presently be generated to have the rotating wave form (RWF)
\begin{eqnarray}
{\bm H}(t)&=&H_0\hat{\bm z}+H_1[\hat{\bm x}\cos(\omega t)+\hat{\bm y}\sin(\omega t)],\label{Hoft}
\end{eqnarray}
over a wide range of angular frequencies $\omega$.  With this form, the time-dependent part of the magnetic field rotates about the time-independent field component.  Usually one has $H_1/H_0<<1$, and for NMR and EPR experiments, one needs to  have $\omega$ variable over at least an order of magnitude around the respective resonant values $\omega_{n0}$ and  $\omega_{e0}$ described in the following.  More complicated periodic time dependencies described by Fourier series can also be presently generated.  

Historically, in early standard magnetic resonance experiments \cite{Majorana,Schwinger,RabiRamseySchwinger,Ramsey,RamseyII}, the Hamiltonian $\hat{H}(t)$ for the interaction of the nuclear spin ${\bm I}$ with the applied magnetic field ${\bm H}(t)$ at the time $t$ has a form similar to
$\hat{H}(t)=\mu_n{\bm I}\cdot[H_0\hat{\bm z}+H_1\hat{\bm x}\cos(\omega t)]$, where $\mu_n$ is the  magnetic moment appropriate for the studied nuclear isotope.

Although the exact solution to this problem has not been published for general $I$,  especially in a form suitable to treat the atomic and molecular interactions of the various nuclear moments with their surrounding electrons, those authors found that one could make great progress by use of magnetic fields of the RWF, in which the oscillatory field component rotates about the constant field component precisely as in Eq. (\ref{Hoft}), the bare, or nominally independent  nuclear Hamiltonian of which is
\begin{eqnarray}
{\hat H}^0_{n}(t)&=&\omega_{n0}I_z+\omega_{n1}[I_x\cos(\omega t)+I_y\sin(\omega t)],\>\>\label{Hn}
\end{eqnarray}
where $\omega_{n0}=\mu_n(1-\sigma)H_0$,
$\omega_{n1}=\mu_n(1-\sigma')H_1$, $\sigma$ and $\sigma'$ are the chemical shifts along and normal to ${\bm z}$,
and usually $|\omega_{n1}/\omega_{n0}|\ll 1$ \cite{Majorana,Schwinger,RabiRamseySchwinger,Ramsey,RamseyII,Schwinger1977,Gottfried,SN3}. We note that the chemical shifts are very weak perturbations, usually measured in parts per million (ppm) due to the particular surrounding electronic environment that has long been used by organic chemists to aid in the identification of the molecular locations of the probed nuclei \cite{KKK1,KKK2,KKSW}.

Early treatments of this  problem in terms of $2I$ Pauli matrices by Majorana, Rabi, Ramsey, and Schwinger were  correct \cite{Majorana,Schwinger,RabiRamseySchwinger,Ramsey,RamseyII,Schwinger1977}, and the derivation for arbitrary ${\bm H}(t)$ and $I$, while  complicated, could  eventually be found if the exact solution for $I=\frac{1}{2}$ were known \cite{Schwinger1977}.

 Gottfried  removed the $t$ dependence of ${\hat H}(t)$ in Eq. (\ref{Hn}) by performing a rotation about the $z$ axis by the $t$-dependent  ``angle'' $-\omega t$, obtaining for general nuclear spin $I$ \cite{Gottfried},
\begin{eqnarray}
|\psi_{G,n}(t)\rangle&=&e^{-iI_z\omega t/\hbar}e^{-i{\hat H}_{{\rm eff},n}t/\hbar}|\psi_{G,n}(0)\rangle,\label{psi1oft}
\end{eqnarray}
where the basis for $|\psi_{G,n}(0)\rangle$ consists of the complete set of eigenstates $|I,m\rangle$ of ${\bm I}^2$ and $I_z$,
\begin{eqnarray}
{\bm I}^2|I,m\rangle&=&\hbar^2I(I+1)|I,m\rangle,\label{I2}\nonumber\\
I_z|I,m\rangle&=&\hbar m|I,m\rangle,\label{Iz}
\end{eqnarray}
and
\begin{eqnarray}
{\hat H}_{{\rm eff},n}&=&(\omega_{n0}-\omega)I_z+\omega_{n1}I_x.\label{Heff}
\end{eqnarray}
Gottfried then correctly wrote the nuclear spin wave function $|\psi_{G,n}(t)\rangle$ (in our notation) as
\begin{eqnarray}
|\psi_{G,n}(t)\rangle&=&e^{-iI_z\omega t/\hbar}e^{-i\hat{\bm n}_n\cdot{\bm I}\Omega_n t/\hbar}|\psi_{G,n}(0)\rangle,
\label{Gottfriedwavefunction}
\end{eqnarray}
where
\begin{eqnarray}
\hat{\bm n}_n&=&\hat{\bm z}\cos(\beta_n)+\hat{\bm x}\sin(\beta_n),\label{nhat}\\
\cos(\beta_n)&=&\frac{\omega_{n0}-\omega}{\Omega_n};\hskip10pt\sin(\beta_n)=\frac{\omega_{n1}}{\Omega_n}\label{sincosbeta}
\end{eqnarray}
and
\begin{eqnarray}
\Omega_n&=&\sqrt{(\omega_{n0}-\omega)^2+\omega_{n1}^2}.\label{Omega}
\end{eqnarray}
Gottfried then wrote that for the probability amplitude for a transition from the single nuclear occupied initial state $|\psi_n(0)\rangle=|I,m\rangle$ to the single occupied final state $|I,m'\rangle$ at the time $t$ is given by
\begin{eqnarray}
\langle I,m'|e^{-iI_z\omega t/\hbar}e^{-i\hat{\bm n}_n\cdot{\bm I}\Omega_n t/\hbar}|I,m\rangle
&=&e^{-im'\omega t}d^{(I)}_{m',m}[\tilde{\beta}_n(t)],\nonumber\\
\label{dmatrix}
\end{eqnarray}
where
\begin{eqnarray}
\sin\Bigl[\frac{1}{2}\tilde{\beta}_n(t)\Bigr]&=&\frac{\omega_{n1}}{\Omega_n}\sin\Bigl(\frac{\Omega_n t}{2}\Bigr),\label{tildebetaoft}
\end{eqnarray}
and therefore that the probability $P_{m',m}(t)$ of a transition from the initial nuclear single-state $|I,m\rangle$ to the  nuclear single-state $|I,m'\rangle$ at the time $t$ is given by
\begin{eqnarray}
P_{m',m}(t)&=&|d^{(I)}_{m',m}[\tilde{\beta}_n(t)]|^2\label{Gottfriedformula}.\end{eqnarray}

 Although  Eqs. (\ref{Gottfriedwavefunction}) to (\ref{Gottfriedformula}) are correct, as shown in Sec. VI, Eqs. (\ref{Gottfriedwavefunction}) and the non-trivial $t$ dependence of (\ref{tildebetaoft}) can be rather cumbersome to employ in perturbation treatments of experimentally relevant Hamiltonians. It is also cumbersome to use it to describe transitions between mixed or entangled states in order to make it useful in quantum computing.   Since the matrix elements of $d^{(I)}_{m',m}(\beta_n)$ are known when $\beta_n$ represents the second Euler angle rotation proportional to  $\langle I,m'|\exp(-i\beta_n I_y/\hbar)|I,m\rangle$ \cite{SN3},  in which the operator $I_y$  is not present in $H_{{\rm eff},n}$, so additional rotations are implicit in Gottfried's expression for $\tilde{\beta_n}(t)$, Eq. (\ref{tildebetaoft}). The equivalence of Gottfried's wave function, Eq. (\ref{Gottfriedwavefunction}),  and the present useful exact wave function derived in Sec. II is shown explicitly for $I=1$ in the Appendix. However, as shown in the following, the present useful exact wave function form is considerably easier to employ in such perturbation expansions.  The present form also allows one to calculate the probability of a transition from a fully general state to another fully general state, which can be useful in quantum computing.
 
 In addition, we note that the second exponential factor in $|\psi_{G,n}(t)\rangle$ in Eq. (\ref{Gottfriedwavefunction}) can be written as $e^{u+v}$ where $u=-it\sin(\beta_n)I_x/\hbar$ and  $u=-it\cos(\beta_n)I_z/\hbar$, where $I_x$ and $I_z$ satisfy the Lie algebra given by Eq. (\ref{commutator}).  One could in principle employ the Campbell-Baker-Hausdorff-Dynkin theorem to rewrite this in an infinite series of powers of $uv$ \cite{CBHD}.  However, such a power series would not only be an infinite series in powers of $t^2I_xI_z$, its  extreme complexity in form would make expressions for transitions from one general state to another extremely complicated.  It would also greatly complicate its use by experimentalists  in measurements of physical properties such as nuclear moments.  
 
  Here  a similar technique to obtain Eqs. (\ref{psi1oft}) and (\ref{Heff}) is used \cite{Majorana,Schwinger,RabiRamseySchwinger,Ramsey,RamseyII,Schwinger1977,Gottfried}, but  an elementary procedure to rewrite the wavefunction in Eq. (\ref{Gottfriedwavefunction}) in terms of products of exponentials of single spin operators valid for arbitrary $I$ and to generalize it to arbitrary $J$ is added \cite{KLK,Kimdissertation,Liudissertation}.  This allows one  to find the useful exact operator form for the general wavefunction $|\psi(t)\rangle$, allowing both the initial state (at $t=0$) and the state at time $t$ to be the most general linear combinations of the substates, and hence to obtain an exact expression for the most general  complete set of eigenstate  occupation amplitudes $C_{m'}(t)$ at an arbitrary time $t\ge0$ based upon their most general complete set of initial  occupation amplitudes $C_m(0)$. Some textbooks \cite{Griffiths,SN3} solved the simplest case $I=\frac{1}{2}$ for one state initially occupied.

In Sec. II,  the derivations of the useful exact  wave functions  for both NMR with arbitrary $I$ and EPR for arbitrary $J$ using  magnetic fields of the form in Eq. (\ref{Hoft}), a combined form useful for accurate perturbation corrections to that magnetic field form are presented in detail. Both are needed to calculate the nuclear moment energies from Eq. (\ref{Delta}).  A fully general expression for the electric quadrupole moment energies is derived, and the exact energies for $I=1, J=\frac{1}{2}$ are found. In Sec. III,  the adiabatic approximation is given for general eigenstates.  In Sec. IV, the occupation amplitudes and probabilities and the resonance condition are given.  In Sec. V,   numerical results for $\frac{1}{2}\le I\le2$ under three different initial conditions are presented.  In Sec. VI, the uses of the exact combined NMR and EPR wave function to treat a variety of Hamiltonians accurately at and near to resonance conditions are presented.  In order to treat the nuclear moment energies, it is necessary to first obtain the analogous useful exact wave function for the electronic angular momentum $J$ states.  Our method can in principle solve exactly for the energies of the entire nuclear moment series in Eq. (\ref{Delta}), which could be crucial in determining the seven nuclear moments of $^{133}$Cs. Finally, in Sec. VII,  the summary and  conclusions are provided.

\section{The  useful exact solution}
\subsection{The useful exact nuclear wave function}

  Starting with $\hat{H}^0_{n}(t)$ in Eq. (\ref{Hn}), one
first rotates $I_x$ and $I_y$ by an angle $\phi$ about the $z$ axis, \cite{Gottfried,Griffiths,BS,Rohlf,SN3,GY}
\begin{eqnarray}
I_x(\phi)&=&e^{iI_z\phi/\hbar}I_xe^{-iI_z\phi/\hbar},\nonumber\\
I_y(\phi)&=&e^{iI_z\phi/\hbar}I_ye^{-iI_z\phi/\hbar}.\label{Ixyofphi}
\end{eqnarray}
Taking the derivatives of these equations with respect to $\phi$, one may write
\begin{eqnarray}
\frac{\partial I_x(\phi)}{\partial\phi}&=&\frac{i}{\hbar}e^{iI_z\phi/\hbar}\left[I_z,I_x\right]e^{-iI_z\phi/\hbar}=-I_y(\phi),\nonumber\\
\frac{\partial I_y(\phi)}{\partial\phi}&=&\frac{i}{\hbar}e^{iI_z\phi/\hbar}\left[I_z,I_y\right]e^{-iI_z\phi/\hbar}=I_x(\phi),\label{derivatives}
\end{eqnarray}
where  Eq. (\ref{commutator}) was used to evaluate the commutators.  By repeating this process, one finds for $i=x,y$,
\begin{eqnarray}
\frac{\partial^2I_i(\phi)}{\partial\phi^2}+I_i(\phi)&=&0,
\end{eqnarray}
which is  the one-dimensional simple harmonic oscillator equation, with solutions $A_i\cos(\phi)+B_i\sin(\phi)$. Imposing the boundary conditions  $I_x(0)=I_x$, $I_y(0)=I_y$, and
  \begin{eqnarray}
  \frac{\partial I_x(\phi)}{\partial\phi}\Bigr|_{\phi=0}&=&-I_y,\nonumber\\
  \frac{\partial I_y(\phi)}{\partial\phi}\Bigr|_{\phi=0}&=&I_x,
  \end{eqnarray}
 from Eqs. (\ref{derivatives}) at $\phi=0$, one easily obtains
\begin{eqnarray}
I_x(\phi)&=&I_x\cos(\phi)-I_y\sin(\phi),\nonumber\\
I_y(\phi)&=&I_y\cos(\phi)+I_x\sin(\phi),\label{Ixyofphianswer}
\end{eqnarray}
as for a classical rotation of the $x$ and $y$ axes by the angle $\phi$ about the $z$ axis \cite{SN3}.
Letting $\phi\rightarrow -\omega t$ \cite{Gottfried}, one obtains $I_x(-\omega t)$ given by
\begin{eqnarray}
I_x\cos(\omega t)+I_y\sin(\omega t)
&=&e^{-iI_z\omega t/\hbar}I_xe^{iI_z\omega t/\hbar}.
\end{eqnarray}
 \begin{figure}
 \center{\includegraphics[width=0.49\textwidth]{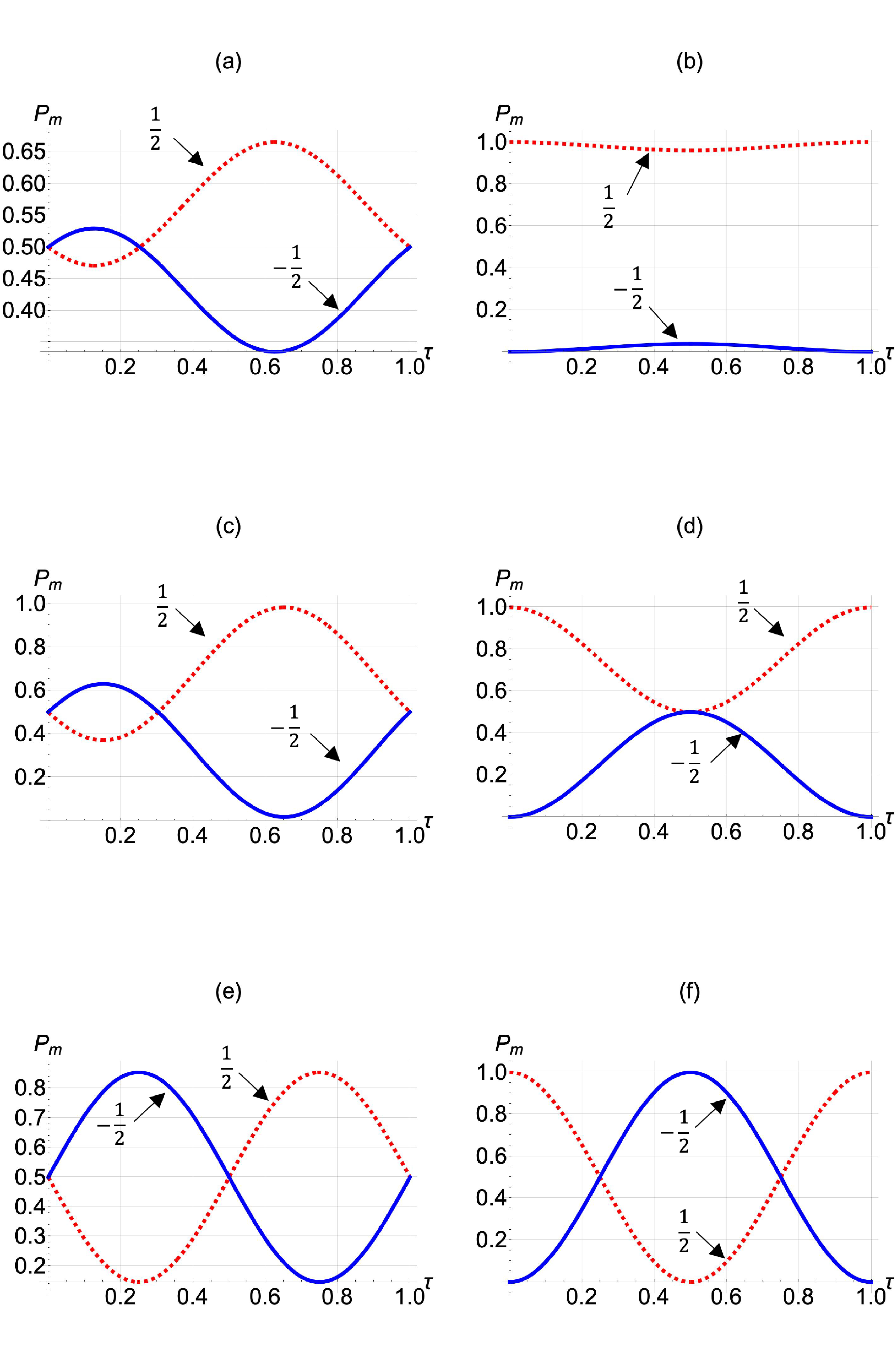}\caption{(Color online) Plots  of $P_{1/2}(\tau)$ (dotted red) and $P_{-1/2}(\tau)$ (solid blue) for $I=\frac{1}{2}$ and $0\le\tau=t/T_{n}\le1$.  In (a), (c), and (e), $C_{1/2}(0)e^{-i\pi/4}=C_{-1/2}(0)=\frac{1}{\sqrt{2}}$.  In (b), (d), and (f), $C_{1/2}(0)=1$ and $C_{-1/2}(0)=0$.  In (a) and (b), 
$\omega/\omega_{n0}=0.95$, in (c) and (d), $\omega/\omega_{n0}=0.99$, and in (e) and (f), $\omega/\omega_{n0}=1$. This figure also applies for $J=\frac{1}{2}$ with $\tau=t/T_e$ and $\omega_{n0}$ and $C_m(0)$ respectively replaced by $\omega_{e0}$ and $D_{\overline{m}}(0)$. }}
\end{figure}

The bare Hamiltonian in Eq. (\ref{Hn}) may then be written as
\begin{eqnarray}
{\hat H}^0_{n}(t)&=&e^{-iI_z\omega t/\hbar}\Bigl(\omega_{n0}I_z+\omega_{n1}I_x\Bigr)e^{iI_z\omega t/\hbar}.
\end{eqnarray}
One may then solve the  Schr{\"o}dinger equation,
\begin{eqnarray}
{\hat H}^0_{n}(t)|\psi^0_{n}(t)\rangle&=&i\hbar\frac{\partial|\psi^0_{n}(t)\rangle}{\partial t}
\end{eqnarray}
by writing
\begin{eqnarray}
|\psi^0_{n}(t)\rangle&=&e^{-iI_z\omega t/\hbar}|\psi^{0'}_{n}(t)\rangle,\label{psitprime}\\
i\hbar\frac{\partial|\psi^0_{n}\rangle}{\partial t}&=&\omega I_ze^{-iI_z\omega t/\hbar}|\psi^{0'}_{n}\rangle+i\hbar e^{-iI_z\omega t/\hbar}\frac{\partial|\psi^{0'}_{n}\rangle}{\partial t},\nonumber
\end{eqnarray}
leading to
\begin{eqnarray}
e^{-iI_z\omega t/\hbar}\Bigl(\omega_{n0}I_z+\omega_{n1}I_x\Bigr)|\psi^{0'}_{n}\rangle&=&e^{-iI_z\omega t/\hbar}\Bigl(\omega I_z|\psi^{0'}_{n}\rangle\nonumber\\
& &+i\hbar\frac{\partial|\psi^{0'}_{n}\rangle}{\partial t}\Bigr),\label{psiprime1}
\end{eqnarray}
or
\begin{eqnarray}
{\hat H}_{{\rm eff},{n}}|\psi^{0'}_{n}\rangle=\Bigl[(\omega_{n0}-\omega)I_z+\omega_{n1}I_x\Bigr]|\psi^{0'}_{n}\rangle&=&i\hbar\frac{\partial|\psi^{0'}_{n}\rangle}{\partial t},\label{psiprime}\nonumber\\
\end{eqnarray}
where to obtain Eq. (\ref{psiprime}), both sides of Eq. (\ref{psiprime1}) were multiplied by $\exp(iI_z\omega t/\hbar)$ on the left, and
where ${\hat H}_{{\rm eff},n}$  is independent of $t$.
Solving Eq. (\ref{psiprime}) for $|\psi'_n(t)\rangle$, one easily finds
\begin{eqnarray}
|\psi^{0'}_{n}(t)\rangle&=&e^{-i{\hat H}_{{\rm eff},n}t/\hbar}|\psi^{0'}_{n}(0)\rangle.\label{psiprime2}
\end{eqnarray}
Combining Eqs. (\ref{psiprime2}) and (\ref{psitprime}) with the initial conditions $|\psi^{0'}_{n}(0)\rangle=|\psi^0_{n}(0)\rangle$,
one obtains Gottfried's Eq. (\ref{psi1oft}) \cite{Gottfried}.
 \begin{figure}
 \center{\includegraphics[width=0.49\textwidth]{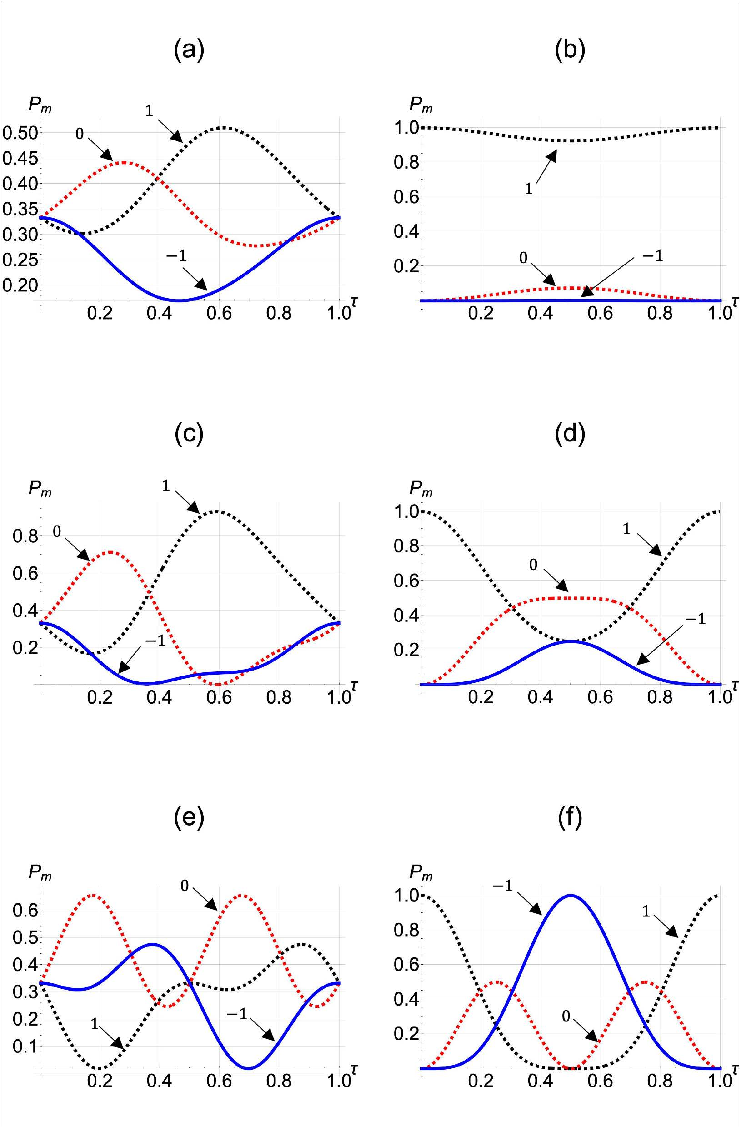}
 \caption{(Color online) Plots  of $P_1(\tau)$ (dotted black), $P_0(\tau)$ (dotted red) and $P_{-1}(\tau)$ (solid blue) for $I=1$ and $0\le\tau=t/T_n\le1$.  In (a), (c), and (e), $C_0(0)=C_{-1}(0)=e^{-i\pi/4}C_1(0)=\frac{1}{\sqrt{3}}$.  In (b), (d), and (f), $C_1(0)=1$ and $C_0(0)=C_{-1}(0)=0$.  In (a) and (b), $\omega/\omega_{n0}=0.95$, in (c) and (d), $\omega/\omega_{n0}=0.99$, and in (e) and (f), $\omega/\omega_{n0}=1$.  This figure also applies for $J=1$ with $\tau=t/T_e$ and $\omega_{n0}$ and $C_m(0)$ respectively replaced by $\omega_{e0}$ and $D_{\overline{m}}(0)$.}}
 \end{figure}

 As Gottfried correctly argued, ${\hat H}_{{\rm eff},n}$ can be diagonalized by  rotating  ${\bm I}$ about the $y$ axis by an angle $\beta_{n}$ so that ${\bm I}(\beta_{n})||\hat{z}$.  In order to determine how the wave function changes, it is useful to diagonalize the second exponential operator factor in the wave function given by Eq. (\ref{psi1oft}), which  Gottfried did not do.   By inserting ${\bm 1}_n=e^{-i\beta_{n}I_y/\hbar}e^{i\beta_{n}I_y/\hbar}$ between and to the right of the two exponential operators in Eq. (\ref{psi1oft}), where  ${\bm 1}_n$ is the identity matrix of rank $2I+1$, one finds that
\begin{eqnarray}
|\psi^0_{n}(t)\rangle&=&e^{-iI_z\omega t/\hbar}e^{-i\beta_{n} I_y/\hbar}{\cal O}(\beta_{n},t)e^{+i\beta_{n} I_y/\hbar}\nonumber\\
& &\times|\psi^0_{n}(0)\rangle,\label{psiobetaoft}
\end{eqnarray}
where
\begin{eqnarray}
{\cal O}(\beta_{n},t)&=&e^{+i\beta_{n} I_y/\hbar}e^{-i{\hat H}_{{\rm eff},n}t/\hbar}e^{-i\beta_{n} I_y/\hbar}.\label{O}
\end{eqnarray}
By defining
\begin{eqnarray}
\hat{H}_{{\rm eff},n}(\beta_n)&=&e^{+i\beta_nI_y/\hbar}{\hat H}_{{\rm eff},n}e^{-i\beta_{n} I_y/\hbar},\label{Heffofbeta}
\end{eqnarray}
one may then evaluate
${\cal O}(\beta_{n},t)$
 by expanding $e^{-i{\hat H}_{{\rm eff},n}/\hbar}$ in Eq. (\ref{O}) in a  power series in ${\hat H}_{{\rm eff},n}$,  inserting ${\bm 1}_n=e^{-i\beta_{n}I_y/\hbar}e^{i\beta_{n}I_y/\hbar}$ between each pair of ${\hat H}_{{\rm eff},n}$ operators, transforming each ${\hat H}_{{\rm eff},n}$ to $\hat{H}_{{\rm eff},n}(\beta_{n})$ by Eq. (\ref{Heffofbeta}), and resumming the series, leading to
\begin{eqnarray}
{\cal O}(\beta_{n},t)&=&e^{-i\hat{H}_{{\rm eff},n}(\beta_{n})t/\hbar}.\label{Obeta}
\end{eqnarray}

Using the same procedure
as in Eqs. (\ref{Ixyofphi})-(\ref{Ixyofphianswer}), the rotated $I_x(\beta_{n})$ and $I_z(\beta_{n})$ in Eq. (\ref{Heffofbeta}) are
\begin{eqnarray}
I_x(\beta_{n})&=&I_x\cos(\beta_{n})+I_z\sin(\beta_{n}),\\
I_z(\beta_{n})&=&I_z\cos(\beta_{n})-I_x\sin(\beta_{n}),
\end{eqnarray}
 so that
\begin{eqnarray}
\tilde{H}_{{\rm eff},n}(\beta_{n})&=&(\omega_{n0}-\omega)[I_z\cos(\beta_{n})-I_x\sin(\beta_{n})]\nonumber\\
& &+\omega_{n1}[I_x\cos(\beta_{n})+I_z\sin(\beta_{n})].
\end{eqnarray}

 \begin{figure}
\center{\includegraphics[width=0.49\textwidth]{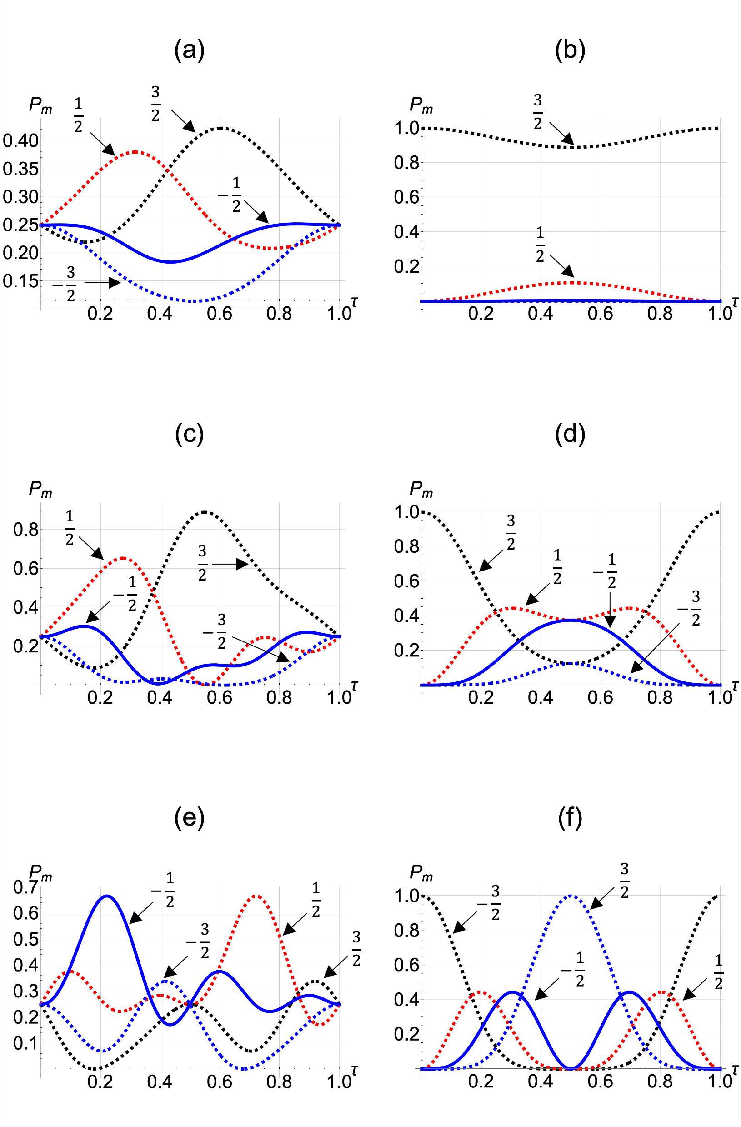} \caption{(Color online) Plots  of $P_{3/2}(\tau)$ (dotted black), $P_{1/2}(\tau)$ (dotted red), $P_{-1/2}(\tau)$ (solid blue), and $P_{-3/2}(\tau)$ (dotted blue) for $I=\frac{3}{2}$ and $0\le\tau=t/T_{n}\le1$.  In (a), (c), and (e), $C_{-\frac{3}{2}}(0)=C_{-\frac{1}{2}}(0)=C_{\frac{1}{2}}(0)=e^{-i\pi/4}C_{\frac{3}{2}}(0)=\frac{1}{2}$.  In (b), (d), and (f), $C_{\frac{3}{2}}(0)=1$ and $C_m(0)=0$ for $m\ne\frac{3}{2}$.  In (a) and (b), $\omega/\omega_{n0}=0.95$, in (c) and (d), $\omega/\omega_{n0}=0.99$, and in (e) and (f), $\omega/\omega_{n0}=1$.  This figure also applies for $J=\frac{3}{2}$ with $\tau=t/T_e$ and $m$, $\omega_{n0}$ and $C_m(0)$ respectively replaced by $\overline{m}$, $\omega_{e0}$, and $D_{\overline{m}}(0)$.}}
\end{figure}

Equations (\ref{sincosbeta}) force the off-diagonal terms to vanish,
leading to
\begin{eqnarray}
\hat{H}_{{\rm eff},n}(\beta_{n})&=&\Omega_{n}I_z,\label{OmegaIz}
\end{eqnarray}
where $\Omega_{n}$ is given by Eq. (\ref{Omega}).
Then from Eqs. (\ref{psiobetaoft}), (\ref{Obeta}), and (\ref{OmegaIz}), one finds
\begin{eqnarray}
|\psi^0_{n}(t)\rangle&=&e^{-iI_z\omega t/\hbar}e^{-i\beta_{n} I_y/\hbar}e^{-iI_z\Omega_{n} t/\hbar}e^{i\beta_{n} I_y/\hbar}|\psi^0_{n}(0)\rangle,\nonumber\\
\label{psioftexact}
\end{eqnarray}
which is  also an  exact but a more useful operator form of $|\psi_{n}(t)\rangle$, where $\beta_{n}$ and $\Omega_{n}$ are respectively given by Eqs. (\ref{sincosbeta}) and (\ref{Omega}).  It is straightforward to prove that  $|\psi_{n}(t)\rangle$ given by Eq. (\ref{psioftexact}) exactly satisfies the Schr{\"o}dinger wave equation,
\begin{eqnarray}
i\hbar\frac{\partial}{\partial t}|\psi^0_{n}(t)\rangle&=&\hat{H}^0_{n}(t)|\psi^0_{n}(t)\rangle,\label{SWE}
\end{eqnarray}
where $\hat{H}^0_{n}(t)$ is given by Eqs. (\ref{Hn}). As discussed  in Sec. IV, we choose to define $\beta_{n}=\sin^{-1}(\omega_{n1}/\Omega_{n})$, which is an even function of $\omega-\omega_{n0}$  that is  independent of $t$.
 \begin{figure}
 \center{\includegraphics[width=0.49\textwidth]{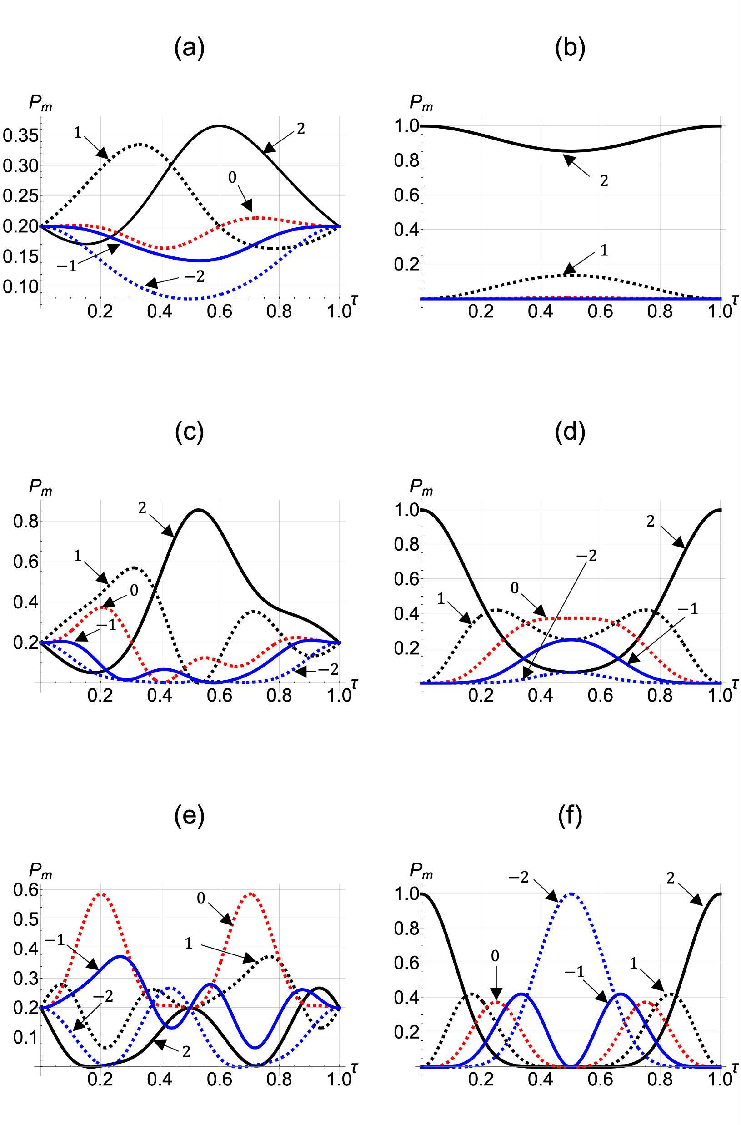}
 \caption{(Color online) Plots  of $P_{2}(\tau)$ (solid black), $P_{1}(\tau)$ (dotted black), $P_{0}(\tau)$ (dotted red), $P_{-1}(\tau)$ (solid blue), and $P_{-2}(\tau)$ (dotted blue) for $I=2$ and $0\le\tau=t/T_{n}\le1$.  In (a), (c), and (e),
 $C_{-2}(0)=C_{-1}(0)=C_{0}(0)=C_{1}(0)=e^{-i\pi/4}C_{2}(0)=\frac{1}{\sqrt{5}}$.  In (b), (d), and (f), $C_{2}(0)=1$ and $C_m(0)=0$ for $m\ne2$.  In (a) and (b), $\omega/\omega_{n0}=0.95$, in (c) and (d), $\omega/\omega_{n0}=0.99$, and in (e) and (f), $\omega/\omega_{n0}=1$.  This figure also applies for $J=2$ with $\tau=t/T_e$ and $m$, $\omega_{n0}$, and $C_m(0)$ respectively replaced by $\overline{m}$, $\omega_{e0}$, and $D_{\overline{m}}(0)$.}}
\end{figure}

Then by writing
\begin{eqnarray}
|\psi^0_{n}(t)\rangle&=&\sum_{m'=-I}^IC_{m'}(t)|I,m'\rangle,\label{psioft}
\end{eqnarray}
taking the inner product $\langle I,m|\psi^0_{n}(t)\rangle$ and using  Eqs. (\ref{psioftexact}) and (\ref{psioft}), the latter both at $t=0$ and at $t > 0$, one obtains
\begin{eqnarray}
C_{m}(t)&=&\sum_{m'=-I}^IC_{m'}(0)\langle I,m|e^{-iI_z\omega t/\hbar}e^{-iI_y\beta_{n}/\hbar}e^{-iI_z\Omega_{n} t/\hbar}\nonumber\\
& &\hskip60pt\times e^{iI_y\beta_{n}/\hbar}|I,m'\rangle,\label{Cmoft}
\end{eqnarray}
where eigenstate orthonormality
\begin{eqnarray}
\langle I,m|I,m'\rangle&=&\delta_{m,m'}\label{orthonormality}
\end{eqnarray}
was employed.
After introducing identity operators such as
\begin{eqnarray}
\sum_{m''=-I}^I|I,m''\rangle\langle I,m''|&=&{\bm 1}_n \label{1n}
\end{eqnarray}
between each of the exponential operators in Eq. (\ref{Cmoft}), and making use of Eqs. (\ref{Iz}) and (\ref{orthonormality}), one finds
\begin{eqnarray}
C_m(t)&=&\sum_{m',m'',m''',m''''=-I}^IC_{m'}(0)\langle I,m|e^{-iI_z\omega t/\hbar}|I,m'''\rangle\nonumber\\
& &\times\langle I,m'''|e^{-i\beta_{n} I_y/\hbar}|I,m''\rangle\nonumber\\
& &\times\langle I,m''|e^{-iI_z\Omega_{n} t/\hbar}|I,m''''\rangle\langle I,m''''|e^{i\beta_{n} I_y/\hbar}|I,m'\rangle\nonumber\\
&=&e^{-im\omega t}\sum_{m',m''=-I}^IC_{m'}(0)e^{-im''\Omega_{n} t}\nonumber\\
& &\times\langle I,m|e^{-i\beta_{n} I_y/\hbar}|I,m''\rangle\langle I,m''|e^{i\beta_{n} I_y/\hbar}|I,m'\rangle\nonumber\\
&=&e^{-im\omega t}\sum_{m',m''=-I}^IC_{m'}(0)e^{-im''\Omega_{n} t}\nonumber\\
& &\hskip70pt\times d_{m,m''}^{(I)}(\beta_{n})d_{m',m''}^{(I)*}(\beta_{n}),\label{amplitude}
\end{eqnarray}
where $d^{(I)}_{m',m}(\beta_{n})$ is the reduced rotation matrix element for the second Euler angle, as derived by Wigner and detailed in textbooks \cite{Gottfried,SN3,GY},

\begin{eqnarray}
d^{(I)}_{m',m}(\beta_{n})&\equiv&\langle I,m'|e^{-i\beta_{n} I_y/\hbar}|I,m\rangle\nonumber\\
&=&\sum_k(-1)^{k-m+m'}\>\>\>\>\>\>\>\nonumber\\
& &\times\frac{\sqrt{(I+m)!(I-m)!(I+m')!(I-m')!}}{(I+m-k)!k!(I-k-m')!(k-m+m')!}\nonumber\\
& &\times\Bigl[\cos(\beta_{n}/2)\Bigr]^{2I-2k+m-m'}\>\>\>\>\>\>\>\>\nonumber\\
& &\times\Bigl[\sin(\beta_{n}/2)\Bigr]^{2k-m+m'},\label{dsm'm}\>\>\>
\end{eqnarray}
where $k$ takes on all values for which none of the arguments of the factorials in the denominator are negative \cite{SN3}.  That is, ${\rm max}(0, m-m')\le k\le {\rm min}(I+m,I-m')$.  In Eq. (\ref{dsm'm}),
\begin{eqnarray}
\sin(\beta_{n}/2)&=&\sqrt{\frac{1-\cos(\beta_{n})}{2}}=\sqrt{\frac{\Omega_{n}+\omega-\omega_{n0}}{2\Omega_{n}}},\label{sinhalfbeta}\\
\cos(\beta_{n}/2)&=&\sqrt{\frac{1+\cos(\beta_{n})}{2}}=\sqrt{\frac{\Omega_{n}+\omega_{n0}-\omega}{2\Omega_{n}}},\label{coshalfbeta}
\end{eqnarray}
 which are both finite and independent of $t$ for any $\omega$.

The $t$ dependence of Eq. (\ref{amplitude}) has  the same form as that of Hall and Klemm \cite{HallKlemm},
who showed it to be valid for $I=\frac{1}{2},1,\frac{3}{2}$, but did not evaluate the constant coefficients. That $T=0$ model was subsequently extended to finite $T$ \cite{Klemm2018}.

Many workers have used the Gottfried wave function for general $I$ or Pauli matrices for $I=\frac{1}{2}$ to treat general $I$ NMR and $N$-level quantum systems exhibiting SU(2) symmetry \cite{Aragone,Rashid,Eberle,BB,CookShore1979,ShoreCook1979,Amirav,Whaley,Prior,Layton,Amirav1991,Irish,Hioe,Petrovic,Zanon}.  Some of those workers used the Gottfried wave function to evaluate the probability of a transition from a single fully-occupied state to another single fully-occupied state, and their figures appear to be identical to ours, even for $I=\frac{9}{2}$ (or $N=10$), as were presented elsewhere \cite{Liudissertation}. In some cases they did it purely by numerical techniques, and in some other cases they used Gottfried's wave function, and in yet some additional cases, they solved large matrices analytically, but did not write down the wave function. But none of them  either wrote Eq. (\ref{psioftexact}) or evaluated analytically the transitions involving two or more initial or final populated states, which is essential in quantum computing.

\subsection{The useful exact bare electronic wave function}

In an NMR experiment, the electrons in an atom or molecule surrounding a nucleus are subject to the same time-dependent applied magnetic field as are the nuclei. Correspondingly, the nuclei in an atom or molecule in an EPR experiment are subject to the same time-dependent applied magnetic field as are the electrons.  In the absence of interactions between the nuclei and the surrounding electrons,  the relevant electronic bare Hamiltonian is

\begin{eqnarray}
\hat{H}^0_{e}(t)&=&\omega_{e0}J_z+\omega_{e1}[J_x\cos(\omega t)+J_y\sin(\omega t)],\>\>\>\>\>\>\label{He}
\end{eqnarray}
where the total electronic angular momentum operator  components $J_i$  also satisfy the standard Lie algebra in Eq. (\ref{commutator}),
and are independent of the nuclear spin operators $I_j$,
but the quantities $\omega_{e0}=\mu_eH_0$ and $\omega_{e1}=\mu_eH_1$ in Eq. (\ref{He}), where
\begin{eqnarray}
\mu_e&=&\frac{ge}{2m_e},\label{mue}
\end{eqnarray} where   $m_e=9.1093837015(28)\times10^{-31}$ {\rm kg} \cite{Heisse} and $e$ are the mass and charge of an electron, and where $(g-2)/2=\frac{\alpha}{2\pi}+\ldots$ \cite{SchwingerQED} includes the complete series of quantum electrodynamics (QED) corrections to the magnetic moment of the electron,  and where $\alpha$ is the fine structure constant. To date, to accuracy $\left(\frac{\alpha}{2\pi}\right)^5$, $(g-2)/2=1159652181.606(11)(12)(229)\times10^{-12}$, where the three uncertainties in parentheses are respectively from the $10^{\rm th}$ order QED calculations, the hadron, and the best measurement of the fine structure constant. \cite{Aoyama} Since $m_p/m_e=1836.152673346(81)$ \cite{Mohr}, recently improved upon by about a factor of 2 in accuracy, \cite{Heisse},   depending upon the particular atom under study, $m_e$ is at least three to five or more orders of magnitude smaller than the mass of the nucleus under study, so that $\omega_{e0}$ and $\omega_{e1}$ are three  to five orders of magnitude larger than and usually opposite in sign to the analogous nuclear quantities $\omega_{n0}$ and $\omega_{n1}$. The magnetic field strengths at the positions of the electrons and the nuclei are assumed to be the same, but the magnetic moments are enormously different, primarily due to the  enormously different electronic and nuclear masses.

Then, we may immediately write down the exact bare electronic wave function
for an applied magnetic field of the form of Eq. (\ref{Hoft}),
\begin{eqnarray}
|\psi^0_{e}(t)\rangle&=&e^{-iJ_z\omega t/\hbar}e^{-i\beta_{e} J_y/\hbar}e^{-i\Omega_{e}J_zt/\hbar}e^{i\beta_{e} J_y/\hbar}|\psi^0_{e}(0)\rangle,\nonumber \\ \label{psiEoft}
\end{eqnarray}
where
\begin{eqnarray}
\sin(\beta_e)&=&\frac{\omega_{e1}}{\Omega_{e}},\hskip20pt\cos(\beta_e)=\frac{\omega_{e0}-\omega}{\Omega_{e}},\label{sincosbetae}\\
\Omega_{e}&=&\sqrt{(\omega-\omega_{e0})^2+\omega_{e1}^2}.\label{Omegae}
\end{eqnarray}
We  note that when $\omega\approx\omega_{n0}$ for nuclear magnetic resonance studies, the applied frequency $\omega$ is three to five orders of magnitude below the electron paramagetic resonance frequency $\omega_{e0}$, depending upon the particular nuclear mass under study.  Correspondingly, in an electron paramagnetic resonance experiment, $\omega\approx\omega_{e0}$, $\omega$ is three to five orders of magnitude larger than $\omega_{n0}$.  As for the nuclear $\beta_n$, we define
\begin{eqnarray}
\beta_e&=&\sin^{-1}\Bigl(\frac{\omega_{e1}}{\Omega_e}\Bigr).
\end{eqnarray}
The bare wave function for the start of the perturbation calculation is
\begin{eqnarray}
|\psi^0_{ne}(t)\rangle&=&|\psi^0_{n}(t)\rangle\otimes|\psi^0_{e}(t)\rangle,\label{barewavefunction}
\end{eqnarray}
where $|\psi^0_{n}(t)\rangle$, $|\psi^0_{e}(t)\rangle$, $\Omega_{n}$, and $\Omega_e$ are respectively given by Eqs. (\ref{psioftexact}), (\ref{psiEoft}), (\ref{Omega}), and (\ref{Omegae}).
Although the $I_i$ and $J_i$ operators operate on orthogonal subspaces, one can formally combine $|\psi^0_{n}(t)\rangle\otimes|\psi^0_{e}(t)\rangle$ by successively commuting the $J_i$ operators in $|\psi^0_{e}(t)\rangle$ to the left, maintaining their respective order.  We then may write the full  atomic wave function as
\begin{eqnarray}
|\psi^0_{ne}(t)\rangle&=&e^{-i\omega t(I_z{\bm 1}_e+{\bm 1}_nJ_z)/\hbar}e^{-i(\beta_nI_y{\bm 1}_e+\beta_e{\bm 1}_nJ_y)/\hbar}\nonumber\\
& &\times e^{-it(\Omega_nI_z{\bm 1}_e+\Omega_e{\bm 1}_nJ_z)/\hbar}\label{atomicwavefunction}\\
& &\times e^{+i(\beta_nI_y{\bm 1}_e+\beta_e{\bm 1}_nJ_y)/\hbar}|\psi^0_{ne}(0)\rangle,\nonumber
\end{eqnarray}
where ${\bm 1}_e$ and ${\bm 1}_n$ are unit matrices of respective ranks $2J+1$ and $2I+1$,
\begin{eqnarray}
|\psi^0_{ne}(0)\rangle&=&\sum_{m=-I}^IC_m(0)|I,m\rangle\otimes\sum_{\overline{m}=-J}^JD_{\overline{m}}(0)|J,\overline{m}\rangle,\label{psineinitial}\nonumber\\
\end{eqnarray}
and
\begin{eqnarray}
\sum_{\overline{m}=-J}^J|D_{\overline{m}}(0)|^2&=&\sum_{m=-I}^I|C_m(0)|^2=1
\end{eqnarray}
to insure normalization.

\section{adiabatic approximation}

In the adiabatic approximation \cite{Griffiths,SN3,GY,Berry}, there are no transitions from one state to another, so that

\begin{eqnarray}
C_m^{\rm ad}(t)&=&C_m(0)e^{-im\omega t}e^{-im\Omega_{n} t}|d^{(I)}_{m,m}(\beta_{n})|^2,\label{adiabatic}
\end{eqnarray} and
\begin{eqnarray}
D_{\overline{m}}^{\rm ad}(t)&=&D_{\overline{m}}(0)e^{-i\overline{m}\omega t}e^{-i\overline{m}\Omega_et}|d^{(I)}_{\overline{m},\overline{m}}(\beta_e)|^2,
\end{eqnarray}

which can be rewritten as \cite{Griffiths,SN3,GY,Berry}
\begin{eqnarray}
C_m^{\rm ad}(t)&=&e^{i\theta_m(t)}e^{i\gamma_m(t)}C_m^{\rm ad}(0),\label{dynamicgeometric}
\end{eqnarray}
 and
\begin{eqnarray}
D_{\overline{m}}^{\rm ad}(t)&=&e^{i\theta_{\overline{m}}(t)}e^{i\gamma_{\overline{m}}(t)}D_{\overline{m}}^{\rm ad}(0).
\end{eqnarray}
To identify the dynamic $\theta_m(t),\theta_{\overline{m}}(t)$ and geometric (or Berry) $\gamma_m(t),\gamma_{\overline{m}}(t)$ phase, it is useful to redefine Eqs. (\ref{Hn}) and (\ref{Omega}) by setting $\omega_{n}\rightarrow\omega_{n0}\cos\alpha_{n}$, $\omega_{n1}\rightarrow\omega_{n0}\sin\alpha_{n}$, $\omega_{e}\rightarrow\omega_{e0}\cos\alpha_{e}$, $\omega_{e1}\rightarrow\omega_{e0}\sin\alpha_{e}$, so that ${\bm H}(t)$ ``sweeps'' along the circular paths at the small angles $\alpha_{n},\alpha_{e}$ about the $z$ axis from the origin to  the surface of a sphere \cite{Griffiths}.
  $\theta_m(t)=-m\omega_{n0}t$ and $\theta_{\overline{m}}(t)=-\overline{m}\omega_{e0}t$ are found by time-integrating  $-E_m/\hbar=-m\omega_{n0}$  and $-E_{\overline{m}}/\hbar=-\overline{m}\omega_{e0}$, where $E_m, E_{\overline{m}}$ are the energies  of the $m^{\rm th}, \overline{m}^{\rm th}$ states obtained from Eq. (\ref{Hn}) with the present settings for $\omega_{n0}$, $\omega_{n1}$, $\omega_{e0}$,  and $\omega_{e1}$ in the $\omega\rightarrow0$ limit \cite{Griffiths,SN3,GY,Berry},
and $\gamma_m(t)$, $\gamma_{\overline{m}}(t)$ are obtained from Eqs. (\ref{adiabatic}), (\ref{dynamicgeometric}) and the above expressions for $\theta_m(t), \theta_{\overline{m}}(t)$ in the low $\omega$ limit far from either resonance, $\omega\ll\omega_{n0},\omega_{e0}.$ \cite{Berry}.
 The geometric phases $\gamma_m(T_0)$ and $\gamma_{\overline{m}}(T_0)$ at $t=T_0=2\pi/\omega$, the long period of the very slow sweeping motion, equal to the time to complete a closed geometrical orbit, are
\begin{eqnarray}
\gamma_m(T_0)&=&-m(\omega+\Omega_{n})T_0-\theta_m(T_0)\nonumber\\
&\approx&2\pi m(\cos\alpha_{n}-1),\label{geometric}
\end{eqnarray}
and
\begin{eqnarray}
\gamma_{\overline{m}}(T_0)&=&-\overline{m}(\omega+\Omega_e)T_0-\theta_{\overline{m}}(T_0)\nonumber\\
&\approx&2\pi \overline{m}(\cos\alpha_e-1),\label{geometrice}
\end{eqnarray}
which apply for any $-I\le m\le I$  and $-J\le\overline{m}\le J$, and agrees  with the standard result for $I,J=\frac{1}{2}$ \cite{Griffiths,SN3,GY,Berry} and with the result for general $-I\le m\le I$  and $-J\le\overline{m}\le J$ \cite{GY}.  Note that for $m=0$, $\theta_0(t)=\gamma_0(T_0)=0$.

\section{Occupation probabilities and resonance}
\subsection{Nuclear spins}

 By interchanging $m$ and $m'$ in Eq. (\ref{Cmoft}),
\begin{eqnarray}
1 = \langle\psi_{n}(t)|\psi_{n}(t)\rangle&=&\sum_{m'=-I}^IP_{m'}(t),\label{completeness}
\end{eqnarray}
where
\begin{eqnarray}
P_{m'}(t)&=&|C_{m'}(t)|^2\nonumber\\
&=&\Bigl|\sum_{m'',m'''=-I}^IC_{m'''}(0)e^{-im''\Omega_{n}t} \nonumber\\
& &\times d^{(I)}_{m',m''}(\beta_{n})d^{(I)*}_{m''',m''}(\beta_{n})\Bigr|^2,\>\>\>\>\>\>\>\> \label{probability}
\end{eqnarray}
is the probability of the occupation of eigenstate $|I,m'\rangle$ at the time $t$.  The sums in Eq. (\ref{probability}) are restricted by  $-I\le m'',m'''\le I$.  Explicitly,  $m'', m'''=-I, -I+1,-I+2,...,I-2,I-1,I$, so for integral $I$, $m'', m'''$  can be any of the integers  from $-I$ to $I$ in the sums, and for half-integral $I$, they can be any  of the half-integral values from $-I$ to $I$ in those sums. In order to compare with Gottfried's result, in the special case in which only the $|I,m\rangle$ state is initially occupied,
\begin{eqnarray}
C_{m'''}(0)&=&\delta_{m,m'''},
\end{eqnarray}
then
\begin{eqnarray}
P_{m'}(t)&=&\Bigl|\sum_{m''=-I}^Ie^{-im''\Omega_{n} t}d^{(I)}_{m',m''}(\beta_{n})d^{(I)*}_{m,m''}(\beta_{n})\Bigr|^2,\>\>\>\>\>\>\>
\end{eqnarray}
which for $I > \frac{1}{2}$ has a qualitatively different mathematical form from Gottfried's also correct  result.  In the above form, there are four $d$ matrices, the argument   $\beta_{n}$  of each is exactly the time-independent, standard second Euler rotation angle, and   the overall $t$ dependence arises only from the elementary exponentials, so that
 $P_{m'}(t)$
is obviously a periodic function of $t$ nominally with period
\begin{eqnarray}
 T_{n}&=&\frac{2\pi}{\Omega_{n}}.\label{T}
   \end{eqnarray}
The $t$ dependence of this result is consistent with those obtained for $I=\frac{1}{2},1,\frac{3}{2}$  by Hall and Klemm \cite{HallKlemm}, although those authors did not provide the details of the time-independent coefficients. For $I=\frac{1}{2}$, this result is also consistent with that of Gottfried \cite{Gottfried}, who obtained the correct result for $I=\frac{1}{2}$ using the Pauli matrix algebra that is much simpler than that of the Lie algebra required for $I\ge1$, and wrote the equivalent result in terms of Eqs. (\ref{Gottfriedwavefunction}), (\ref{dmatrix}), and (\ref{tildebetaoft}). Note that in the above expression, the $t$ dependence of $P_{m'}(t)$ arises from the summation over the $m''$ state indices that for $I > \frac{1}{2}$ can be different from both the initial $m$  and the final $m'$ state indices.  In Gottfried's result with only two $d$ matrix factors, the $t$ dependence is implicitly given by Eq. (\ref{tildebetaoft}).

  Resonance occurs for general $I$ when $\omega=\omega_{n0}$, for which  $\beta_{n}=\frac{\pi}{2}$ in Eqs. (\ref{sinhalfbeta}) and (\ref{coshalfbeta}) and the $d^{(I)}_{m',m}(\frac{\pi}{2})$ in Eq. (\ref{dsm'm}) are independent of $\omega,\omega_{n0},$ and $\omega_{n1}$.

$|\psi_{RWA,n}(t)\rangle$ in Eq. (\ref{psioftexact}) and $C_m(t)$ in Eq. (\ref{amplitude}) have overall  phase factors proportional to $\omega$ that are different for $\omega=\omega_{0}\pm\delta\omega$.  However,
 by defining $\beta_{n}=\sin^{-1}(\omega_{n1}/\Omega_{n})$ from the left expression in Eq. (\ref{sincosbeta}), and since $\Omega_{n}$ given by Eq. (\ref{Omega}), $\beta_n$ is an even function of $\omega-\omega_{n0}$ and $P_m(t)$ is also an even function of $\omega-\omega_{n0}$, which were verified numerically using Eqs. (\ref{dsm'm}) to (\ref{coshalfbeta}).  It has been well established that for $I=\frac{1}{2}$, the resonance curves of $P_{\pm\frac{1}{2}}(t)$ have Lorentzian forms \cite{Gottfried,Griffiths,SN3,GY}.  However, for general $I$ values, especially for integral $I$, the consequences of resonance depend more strongly upon the initial conditions than they do for the $I=\frac{1}{2}$ case presented in existing textbooks \cite{Gottfried,Griffiths,SN3,GY}.

\subsection{Electronic total angular momentum states}
The probability of occupation of the $-J\le\overline{m}'\le J$ electronic state is exactly analogous to that of the $m'$th nuclear state in the previous subsection, except that $I$ is replaced by $J$, and $\beta_n$, $\Omega_n$, and $T_n$  are respectively replaced by $\beta_e$, $\Omega_e$, and $T_e=2\pi/\Omega_e$. For example, we have
\begin{eqnarray}
P_{\overline{m}'}(t)&=&|D_{\overline{m}'}(t)|^2\nonumber\\
&=&\Bigl|\sum_{\overline{m}'',\overline{m}'''=-J}^JD_{\overline{m}'''}(0)e^{-i\overline{m}''\Omega_et} \nonumber\\
& &\times d^{(I)}_{\overline{m}',\overline{m}''}(\beta_e)d^{(I)*}_{\overline{m}''',\overline{m}''}(\beta_e)\Bigr|^2.\>\>\>\>\>\>\>\> \label{probabilitye}
\end{eqnarray}

\section{Numerical results for $\frac{1}{2}\le I,J\le2$}
\subsection{Time-dependent occupation probability functions}

To present some of the more interesting features of magnetic resonance for general $I$  or $J$, the experimentally relevant ratios $\omega_{n1}/\omega_{n0}=0.01$  and $\omega_{e1}/\omega_{e0}=0.01$ in Figs. 1 - 4 were chosen.
In Figs. 1 - 4,  plots  of $P_m(\tau)$ for either $0\le\tau=t/T_{n}\le1$ or $0\le\tau=t/T_e\le1$, where $T_{n}$ is given by Eq. (\ref{T})  and $T_e=2\pi/\Omega_e$, and the $m,\overline{m}$ substates for all $I, J$ values from $\frac{1}{2}$ to  2  with two different initial conditions are presented.  In each of these figures, the top, middle, and bottom figure pairs are for $\omega/\omega_{n0}$  or $\omega/\omega_{e0}$=0.95, 0.99, and 1, which are respectively in the tail, at the half maximum, and at resonance, or the maximum of the assumed Lorentzian lineshapes. The results for $\omega/\omega_{n0}$  or $\omega/\omega_{e0}$ =0.95, 0.99 were also obtained  respectively for $\omega/\omega_{n0}$  or $\omega/\omega_{e0}$=1.05, 1.01. In each row, the left figures assume that all $P_m(0)$ are equal, but  respectively $C_I(0)$  and $D_J(0)$ have phase differences of $\pi/4$ from all of the other $C_m(0)$  or $D_{\overline{m}}(0)$ values, and the right figures are for  $C_{I}(0)=1$ or $D_J(0)=1$ and the other $C_m(0)=0$  or $D_{\overline{m}}(0)=0$.  In each of figures (b), (d), and (f), the $P_m(\tau)$  or $P_{\overline{m}}(\tau)$ curves all  exhibit even reflection symmetry about $\tau=\frac{1}{2}$, but in (a), (c), and (e), the $P_m(\tau)$  or $P_{\overline{m}}(\tau)$ curves have no such symmetry, although at resonance, the $P_m(\tau)$  or $P_{\overline{m}}(\tau)$ curves in each (e) figure are all equal at  $\tau=0,\frac{1}{2},1$, and the $P_m(\tau)$ and $P_{-m}(\tau)$  or $P_{\overline{m}}(\tau)$ and $P_{-\overline{m}}(\tau)$ curves in each (e) and (f) figure are identical but out of phase by $T_{n}/2$  or $T_e/2$.  For integral $I$ and $m=0$  or integral $J$ and $\overline{m}=0$, the resonant period of $P_0(\tau)$ abruptly changes to $T_{n}/2$  or $T_e/2$  for both initial conditions shown.  Analogous cases for $\frac{5}{2}\le I,J\le\frac{9}{2}$ are shown elsewhere \cite{Liudissertation}.

The simplest cases for $I=\frac{1}{2}$ or $J=\frac{1}{2}$ are shown in Fig. 1.  For the equal initial probability condition, Figs. 1(a) and 1(c) show that the maximal values of $P_{\frac{1}{2}}(\tau)$ are larger than those of $P_{-\frac{1}{2}}(\tau)$.  But at resonance, this changes dramatically. Figure 1(e) shows that the $P_{\pm\frac{1}{2}}(\tau)$ curves are identical, except for a phase difference of $T_n/2$  or $T_e/2$.  In addition, the periodicity of their points of equal probability is also $T_n/2$  or $T_e/2$.  However, for the second initial condition $C_{\frac{1}{2}}(0)=1$  or $D_{\frac{1}{2}}(0)=1$, Figs. 1(b) and 1(d) show that as resonance is approached, the maximal values of $P_{-\frac{1}{2}}(\tau)$ increase, but are respectively smaller than the maximal values of $P_{\frac{1}{2}}(\tau)$.  However, at resonance, Fig. 1(f) shows that the $P_{\pm\frac{1}{2}}(\tau)$ are equivalent to each other, identically varying from 0 to 1, except for a phase difference of $T_n/2$  or $T_e/2$.   Although this problem has been solved in textbooks, the authors all assumed one of the $C_{\pm\frac{1}{2}}(0)=0$  or $D_{\pm\frac{1}{2}}(0)=0$, as in Figs. 1(b), 1(d), and 1(f), so the infinite phase difference degeneracy, one example of which is displayed in Figs. 1(a), 1(c), and 1(e), could be ignored \cite{Gottfried,Griffiths,SN3,GY}.

The $P_m(\tau)$  curves for $I=1$  are shown in Fig. 2. In Figs. 2(a) and 2(c), it can be seen that when all the
$P_m(0)$ are equal, but $C_{1}(0)$ has a phase shift from the other $C_m(0)$, the $P_m(\tau)<1$ curves are all different, except at $\tau=0, 1$. In comparing Figs. 2(a) and 2(c), it is evident that for the particular $P_m(0)$ case, the maximal values of $P_m(\tau)$ decrease monotonically with $m$, just as for $I=\frac{1}{2}$, as shown in  Fig. 1. However, this changes dramatically at resonance.  In Fig. 2(e), $P_0(\tau)$ has the largest maximum value,  the $P_{\pm1}(\tau)$ are equivalent but out of phase by $T_n/2$ with each other, and the periodicity of the points of equality of all three states is also $T_n/2$.  For the other initial condition, $C_1(0)=1$, by comparing Figs. 2(b) and 2(d), as resonance is approached, the $P_m(\tau)$ for the initially unoccupied states grow, but the maximal values of the $P_m(\tau)$ continue to  decrease monotonically with $m$.  But then  at resonance, Fig. 2(f) shows that the period of $P_0(\tau)$  changes abruptly from $T_n$ to $T_n/2$.   In Fig. 2(f), the equivalent $P_{\pm1}(\tau)$ alternate between 0 and 1, and are out of phase by $T_n/2$ with each other, and the maximal value of $P_0(\tau)$ is $\frac{1}{2}$.  This figure also applies for $J=1$ with $m$, $P_m(\tau)$, $C_m(0)$ and $T_n$ respectively replaced  by $\overline{m}$, $P_{\overline{m}}(\tau)$, $D_{\overline{m}}(0)$, and $T_e$.

In Fig. 3, the analogous $P_m(\tau)$ curves are shown for $I=\frac{3}{2}$.  For the equal probability case with $C_{\frac{3}{2}}(0)$ having a phase shift from the other $C_m(0)$, the $P_m(\tau)<1$ curves are all differently periodic functions with period $T_{n}$, except at resonance.  By comparing Figs. 3(a) and 3(c), as resonance is approached, the maximal values of $P_m(\tau)$ for this initial condition are rank-ordered (or decrease monotonically) with $m$.  But at resonance,  Fig. 3(e) shows that the $P_{\pm\frac{1}{2}}(\tau)$  have the largest maxima and are out of phase with each other by $T_n/2$, as are the generally smaller $P_{\pm\frac{3}{2}}(\tau)$ maxima. From Figs. 3(b) and 3(d), it is apparent that as resonance is approached, the $P_m(\tau)$ maxima increase monotonically in the order of decreasing $m$. However, this changes dramatically at resonance. For both boundary conditions at resonance, $P_{\pm\frac{3}{2}}(\tau)$ and $P_{\pm\frac{1}{2}}(\tau)$ are respectively equivalent functions out of phase with each other by $T_n/2$,  and the periodicity of the points of equality of the four curves is again $T_n/2$.  But at resonance in Fig. 3(f), the equivalent $P_{\pm\frac{3}{2}}(\tau)$ alternate between 0 and 1, and are out of phase by $T_n/2$ with each other, and the usually smaller equivalent $P_{\pm\frac{1}{2}}(\tau)$ alternate between 0 and $\frac{3}{8}$ and are also out of phase with each other by $T_n/2$.    This figure also applies for $J=\frac{3}{2}$ with $m$, $P_m(\tau)$, $C_m(0)$ and $T_n$ replaced  by $\overline{m}$, $P_{\overline{m}}(\tau)$, $D_{\overline{m}}(0)$, and $T_e$, respectively.

In Fig. 4, the analogous $P_m(\tau)$ curves for $I=2$ are shown.  For the cases of the occupations of each state being initially equally probable, Figs. 4(a) and 4(c) show that away from resonance, the maximal values of $P_m(\tau)$ either decrease monotonically with $m$ or saturate  at their initial values.     However, at resonance this changes dramatically.    Fig. 4(e) shows that at resonance, $P_0(\tau)\le 1$ has the largest maximal value. The $\tau$ points of equal probability are again 0, $\frac{1}{2}$ and 1. The $P_{\pm m}(\tau)$ for $m\ne0$ are equivalent curves that are $T_n/2$ out of phase with each other.  In comparing the right column Figs. 4(b) and 4(d), it is clear that in each of these integer $I$ cases, as resonance is approached, the increase in the maximal values of $P_m(\tau)$ occurs monotonically with decreasing $m$, until at resonance, dramatic changes occur. At resonance, Fig. 4(f) also shows that for $C_I(0)=1$,  the  largest maximal $0\le P_{\pm I}(\tau)\le 1$ are equivalent functions that are out of phase by $T_n/2$ with each other, and all of the $P_{\pm m}(\tau)$ for $m\ne0$ are equivalent to each other but out of phase by $T_n/2$ with each other.    The order of their maxima has dramatically changed from a monotonic decrease with $m$ as resonance is approached, to a monotonic decrease with decreasing $|m|$ at resonance, at which the period of $P_0(\tau)$ abruptly changes to $T_n/2$.  This figure also applies for $J=2$ with $m$, $P_m(\tau)$, $C_m(0)$ and $T_n$ replaced  by $\overline{m}$, $P_{\overline{m}}(\tau)$, $D_{\overline{m}}(0)$, and $T_e$, respectively.

 \begin{figure}
 \center{\includegraphics[width=0.49\textwidth]{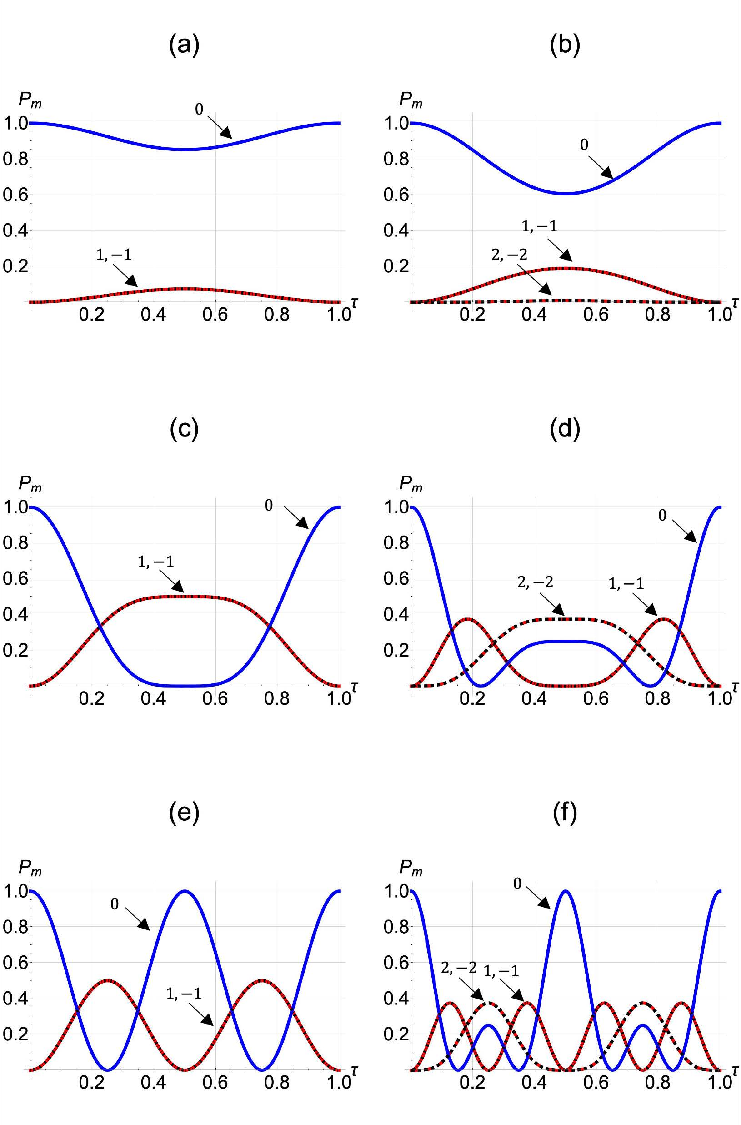}
 \caption{(Color online) Plots of $P_m(\tau)$ for $0\le\tau=t/T_n\le 1$. (a), (c), (e): $I=1$. (b), (d), (f): $I=2$.  In both cases $C_0(0)=1$.  Plots of $P_0(\tau)$ (solid blue), $P_1(\tau)$ (solid red), and $P_{-1}(\tau)$ (dot-dashed black).  For $I=2$, $P_2(\tau)$ (dashed red), and $P_{-2}(\tau)$ (dotted black). In (a) and (b), $\omega/\omega_{n0}=0.95$, in (c) and (d), $\omega/\omega_{n0}=0.99$, and in (e) and (f), $\omega/\omega_{n0}=1$.  This figure also applies for $J=1,2$ with $\tau=t/T_e$ and $m$, $\omega_{n0}$, and $C_m(0)$ respectively replaced by $\overline{m}$, $\omega_{e0}$, and $D_{\overline{m}}(0)$. }}
\end{figure}

 There are an infinite number of possible initial conditions, even for $I,J=\frac{1}{2}$, since the only condition upon the initial eigenstate occupation probabilities is Eq. (\ref{completeness}), so that the phases of each of the $C_m(0)$  or $D_{\overline{m}}(0)$ are arbitrary.  No textbooks solved any case other than $I=\frac{1}{2}$, and usually set one of the $C_{\pm\frac{1}{2}}(0)=0$ or $D_{\pm\frac{1}{2}}(0)=0$ \cite{Gottfried,Griffiths,SN3}, so that the relative phases of the two states were irrelevant. But as seen in the left panels of Fig. 1, the relative phases of the $C_{\pm\frac{1}{2}}(0)$ or $D_{\pm\frac{1}{2}}(0)=0$ are indeed relevant and measurable.
 \begin{figure}
 \center{\includegraphics[width=0.49\textwidth]{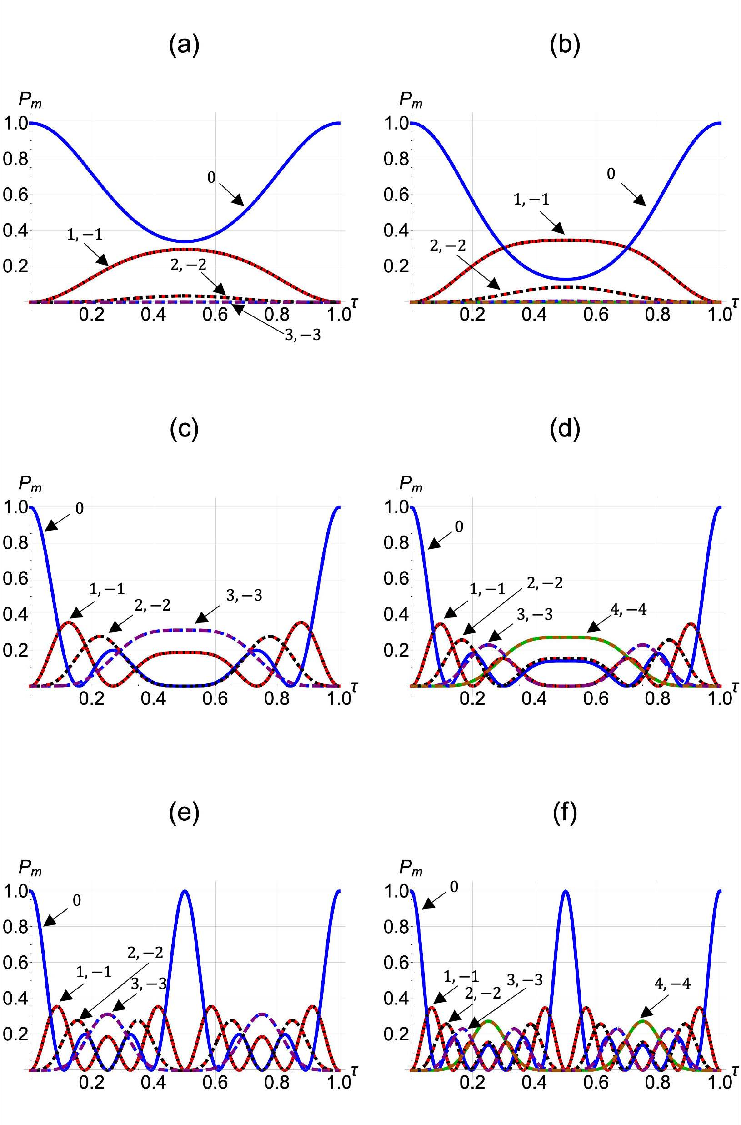}
 \caption{(Color online)  Plots of $P_m(\tau)$ for $0\le\tau=t/T_n\le 1$. (a), (c), (e): $I=3$. (b), (d), (f): $I=4$.  In both cases $C_0(0)=1$.  Plots of $P_0(\tau)$ (solid blue), $P_1(\tau)$ (solid red), and $P_{-1}(\tau)$ (dot-dashed black), $P_2(\tau)$ (dashed red),  $P_{-2}(\tau)$ (dotted black), $P_3(\tau)$ (dashed blue), $P_{-3}(\tau)$ (dotted red). For $I=4$, $P_4(\tau)$ (solid green) and $P_{-4}(\tau)$ (dashed orange).  In (a) and (b), $\omega/\omega_{n0}=0.95$, in (c) and (d), $\omega/\omega_{n0}=0.99$, and in (e) and (f), $\omega/\omega_{n0}=1$. This figure also applies for $J=3,4$ with $\tau=t/T_e$ and $m$,
$\omega_{n0}$, and $C_m(0)$ respectively replaced by $\overline{m}$, $\omega_{e0}$, and $D_{\overline{m}}(0)$. }}
\end{figure}

 It is interesting to show the results for $C_0(0)=1$  or $D_0(0)=1$ in the cases of integral $I$  or $J$.  In Figs. 5 and 6, we show the four cases of $C_0(0)=1$  or $D_0(0)=1$ for $I=1, 2, 3$, and 4.  Figures 5(a), 5(c), and 5(e) show the analogous cases with $C_0(0)=1$ or $D_0(0)=1$ for $I,J=1$, Figs. 5(b), 5(d), and 5(f) are for $C_0(0)=1$  or $D_0(0)=1$ and $I,J=2$, and the analogous cases for $I,J=3$ and 4 are in Figs. 6(a), 6(c), 6(e), and 6(b), 6(d), and 6(f).  In each of these cases, regardless of the proximity to resonance, the $P_{\pm m}(\tau)$  or $P_{\pm \overline{m}}(\tau)$
 are exactly the same for all $m, \overline{m}\ne0$ for integral $I,J$ from 1 to 4. On the approach to resonance, the amplitudes of $P_{\pm m}(\tau)$  or $P_{\pm \overline{m}}(\tau)$ increase monotonically with $|m|$  or $|\overline{m}|$.   But at resonance, the behaviors of the maxima of $P_{\pm m}(\tau)$  or $P_{\pm \overline{m}}(\tau)$ are generally non-monotonic with  $|m|$  or $|\overline{m}|$.  One special case is shown in Fig. 6(f).  In this case, the identical resonant $P_{\pm 1}(\tau)$ are periodic with period $T_{n,e}/4$.  All  of the other curves have identical resonant $P_{\pm m}(\tau)$ that are periodic with period $T_{n,e}/2$.  Also, the order of the decreasing resonant maxima for $I,J=1$   is  ${\rm max}[P_0(\tau)]=1>{\rm max}[P_{\pm 1}(\tau)]=\frac{1}{2}$, for $I,J=2$ it is ${\rm max}[P_0(\tau)]=1>{\rm max}[P_{\pm 1}(\tau)]={\rm max}[P_{\pm2}(\tau)]=\frac{3}{8}$, for $I,J=3$ it is ${\rm max}[P_{ 0}(\tau)]=1>{\rm max}[P_{\pm 1}(\tau)]>{\rm max}[P_{\pm 3}(\tau)]>{\rm max}[P_{\pm 2}(\tau)]$, and for $I,J=4$ it is ${\rm max}[P_{ 0}(\tau)]=1>{\rm max}[P_{\pm 1}(\tau)]>{\rm max}[P_{\pm 4}(\tau)]>{\rm max}[P_{\pm 3}(\tau)]>{\rm max}[P_{\pm 2}(\tau)]$, which are non-monotonic.

\subsection{Period-averaged occupation probabilities near to resonance}

For clarity near to resonance, it is useful to compare the time averages of the occupation probabilities given by

\begin{eqnarray}
\overline{P}_{m,\overline{m}}&=&\frac{1}{T_{n,e}}\int_0^{T_{n,e}}P_{m,\overline{m}}(t)dt
\end{eqnarray}
in the vicinity of $\omega/\omega_{n,e0}\approx1$.  In Figs. 7-10, we present four near-resonant cases for each $I,J$ value satisfying $\frac{1}{2}\le I,J\le 2$.  For each figure, subfigure (a)  contains  plots of the $\overline{P}_{m,\overline{m}}$ for the cases of equal $C_m(0)=\frac{1}{\sqrt{2I+1}}$ or $D_{\overline{m}}(0)=\frac{1}{\sqrt{2J+1}}$, subfigure (b) contains  plots of the $\overline{P}_{m,\overline{m}}$ for the cases $C_I(0)=e^{-i(\pi/2+0.2)}/\sqrt{2I+1}$ for $m=I$ or $D_J(0)=e^{-i(\pi/2+0.2)}/\sqrt{2J+1}$ for $\overline{m}=J$ and  $C_m(0)=\frac{1}{\sqrt{2I+1}}$ for $m\ne I$ or $D_{\overline{m}}(0)=\frac{1}{\sqrt{2J+1}}$ for $\overline{m}\ne J$, and subfigure (c) contains  plots of the $\overline{P}_{m,\overline{m}}$ for the cases $C_m(0)=e^{i\phi_m}/\sqrt{2I+1}$ or $D_{\overline{m}}(0)=e^{i\phi_{\overline{m}}}/\sqrt{2J+1}$, where each of the $\phi_m$ or $\phi_{\overline{m}}$ are random numbers between 0 and $2\pi$.  In each subfigure (d), the plots the $\overline{P}_{m,\overline{m}}$ for the initial case $C_I(0)=1$ or $D_J(0)=1$ and the $C_m(0)=0$ for $m\ne I$  or $D_{\overline{m}}(0)=0$ for $\overline{m}\ne J$ are shown.  In Fig. 7 the $\overline{P}_{m,\overline{m}}$ are plotted over the range $0.95\le\omega/\omega_{n,e0}\le1.05$, and in Figs. 8- 10, $\overline{P}_{m,\overline{m}}$ is plotted over the range $0.99\le\omega/\omega_{n,e0}\le1.01$.  Additional figures for $\frac{5}{2}\le I,J\le\frac{9}{2}$ are shown in the Supplementary Material\cite{SM}.

In Fig. 7,  four cases for $I,J=\frac{1}{2}$ are presented.  In Fig. 7(a), the initial condition $C_{\frac{1}{2}}(0)=C_{-\frac{1}{2}}(0)$ or $D_{\frac{1}{2}}(0)=D_{-\frac{1}{2}}(0)$  leads near resonance to $\overline{P}_{\frac{1}{2}}(\omega/\omega_{n,e0})\ge\overline{P}_{-\frac{1}{2}}(\omega/\omega_{n,e0})$, the equality only holding at resonance, $\omega=\omega_{n,e0}$.  For $C_{\frac{1}{2}}(0)=e^{-i(\pi/2+0.2)}C_{-\frac{1}{2}}(0)$  or $D_{\frac{1}{2}}(0)=e^{-i(\pi/2+0.2)}D_{-\frac{1}{2}}(0)$, the near-resonant $\overline{P}_{\pm\frac{1}{2}}(\omega/\omega_{n,e0})$ curves shown in Fig. 7(b) are nearly inverted with respect to the curves shown in Fig. 7(a), although a close examination of the vertical scales reveals that the inverted curves in Fig. 7(b) are numerically distinct from those in Fig. 7(a), except at resonance.  With randomly chosen different initial $C_{m}(0)$ phases $\phi_m $  or $D_{\overline{m}}(0)$ phases $\phi_{\overline{m}}$, the curves shown in Fig. 7(c) are nearly identical to those shown in Fig. 7(b), but the magnitudes of the curves are again different in the two figures, except exactly at resonance.  In Fig. 7(d) for $C_{-\frac{1}{2}}=0$ or $D_{-\frac{1}{2}}=0$, $\overline{P}_{\frac{1}{2}}\gg\overline{P}_{-\frac{1}{2}}$, except very close to resonance,  at which $\overline{P}_{\frac{1}{2}}=\overline{P}_{-\frac{1}{2}}$.

The analogous cases for $I,J=1$ are shown in Fig. 8.  For all of the $C_m(0)=\frac{1}{\sqrt{3}}$, $\overline{P}_1(\omega/\omega_{n,e0})>\overline{P}_0(\omega/\omega_{n,e0})>\overline{P}_{-1}(\omega/\omega_{n,e0})$, except very near to resonance, as shown in Fig. 8(a). At resonance, $\overline{P}_0(1)>\overline{P}_1(1)=\overline{P}_{-1}(1)$.
  However, when either $C_1(0)=e^{-i(\pi/2+0.2)}\frac{1}{\sqrt{3}}$ and  $C_0(0)=C_{-1}(0)=\frac{1}{\sqrt{3}}$, or $D_1(0)=e^{-i(\pi/2+0.2)}\frac{1}{\sqrt{3}}$ and  $D_0(0)=D_{-1}(0)=\frac{1}{\sqrt{3}}$ away from resonance, $\overline{P}_0(\omega/\omega_{n,e0})>\overline{P}_1(\omega/\omega_{n,e0})>\overline{P}_{-1}(\omega/\omega_{n,e0})$, except very near to resonance, at which $\overline{P}_1(1)=\overline{P}_{-1}(1)>\overline{P}_0(1)$, as shown in Fig. 8(b).  For the particular initial condition example of equal amplitudes and random phases of  either the $C_m(0)$ or $D_{\overline{m}}(0)$ shown in Fig. 8(c), sufficiently far from resonance, $\overline{P}_0(\omega/\omega_{n,e0})>\overline{P}_{-1}(\omega/\omega_{n,e0})>\overline{P}_1(\omega/\omega_{n,e0})$, and at resonance,  $\overline{P}_0(1)>\overline{P}_1(1)=\overline{P}_{-1}(1)$.  For the case of either $C_1(0)=1$  or $D_1(0)$ shown in Fig. 8(d), $\overline{P}_1(\omega/\omega_{n,e0})\gg\overline{P}_0(\omega/\omega_{n,e0})\gg\overline{P}_1(\omega/\omega_{n,e0})$, except very close to resonance, at which $\overline{P}_1(1)=\overline{P}_{-1}(1)>\overline{P}_0(1)$. We note that the functional dependencies of  either the $\overline{P}_m(\omega/\omega_{n0})$  or the $\overline{P}_{\overline{m}}(\omega/\omega_{e0})$ are distinctly different in Fig. 8(d) from the respective forms in Figs. 8(a) to 8(c).  In other words, the same figures and statements for $J=1$ apply by replacing $C_m(0)$ with $D_{\overline{m}}(0)$ and $\omega_{n0}$ with $\omega_{e0}$.

Figure 9 presents the corresponding figures for $I,J=\frac{3}{2}$. At resonance, Figs. 9(a), 9(b), and 9(c) all show $\overline{P}_{\frac{1}{2}}(1)=\overline{P}_{-\frac{1}{2}}(1)>\overline{P}_{\frac{3}{2}}(1)=\overline{P}_{-\frac{3}{2}}(1)$.  Although in Fig. 9(a), the behavior at $\omega/\omega_{n,e0}=1\pm0.01$ (the edges of the figure) is monotonic in $m$  or $\overline{m}$ in $\overline{P}_{m,\overline{m}}(\omega/\omega_{n,e0})$ with $\overline{P}_{\frac{3}{2}}(\omega/\omega_{n,e0}) > \overline{P}_{\frac{1}{2}}(\omega/\omega_{n,e0}) > \overline{P}_{-\frac{1}{2}}(\omega/\omega_{n,e0}) > \overline{P}_{-\frac{3}{2}}(\omega/\omega_{n,e0})$, that monotonicity in $m$  or $\overline{m}$ is not present in Figs. 9(b) and 9(c). Thus the phases of the initial amplitudes modify the non-resonant results substantially. In addition, Fig. 9(d) shows that either for $C_{\frac{3}{2}}(0)=1$  or for $D_{\frac{3}{2}}(0)=1$, sufficiently away from resonance, $\overline{P}_{\frac{3}{2}}(\omega/\omega_{n,e0}) > \overline{P}_{\frac{1}{2}}(\omega/\omega_{n,e0}) > \overline{P}_{-\frac{1}{2}}(\omega/\omega_{n,e0}) > \overline{P}_{-\frac{3}{2}}(\omega/\omega_{n,e0})$, which is also monotonic in  either $m$  or $\overline{m}$, but at resonance, $\overline{P}_{\frac{3}{2}}(1)=\overline{P}_{-\frac{3}{2}}(1)>\overline{P}_{\frac{1}{2}}(1)=\overline{P}_{-\frac{1}{2}}(1)$, opposite to the resonant behavior shown in Fig. 9(a).

The corresponding figures for $I,J=2$ are presented in Fig. 10.  At resonance, $\overline{P}_0(1)>\overline{P}_1(1)=\overline{P}_{-1}(1)>\overline{P}_2(1)=\overline{P}_{-2}(1)$ in Figs. 10(a), and 10(b), but that order is not obeyed at resonance in Fig. 10(c), unlike the behavior in Fig. 9(c) for $I,J=\frac{3}{2}$.  As for the monotonicity in  either $m$  or $\overline{m}$ at $\omega/\omega_{n,e0}=1\pm0.01$, Figs. 10(a) and  10(d)  both display $\overline{P}_2(\omega/\omega_{n,e0}) > \overline{P}_1(\omega/\omega_{n,e0}) > \overline{P}_0(\omega/\omega_{n,e0}) > \overline{P}_{-1}(\omega/\omega_{n,e0}) > \overline{P}_{-2}(\omega/\omega_{n,e0})$, but as in Figs. 9(b) and 9(c), that monotonicity at those values away from resonance are not present in Figs. 10(b) and 10(c).  In addition, at resonance, Fig. 10(d) shows for  either $C_2(0)=1$  or $D_2(0)$ that $\overline{P}_2(1)=\overline{P}_{-2}(1)>\overline{P}_1(1)=\overline{P}_{-1}(1)>\overline{P}_0(1)$, in analogy with Fig. 5(d).

\begin{figure}
\center{\includegraphics[width=0.49\textwidth]{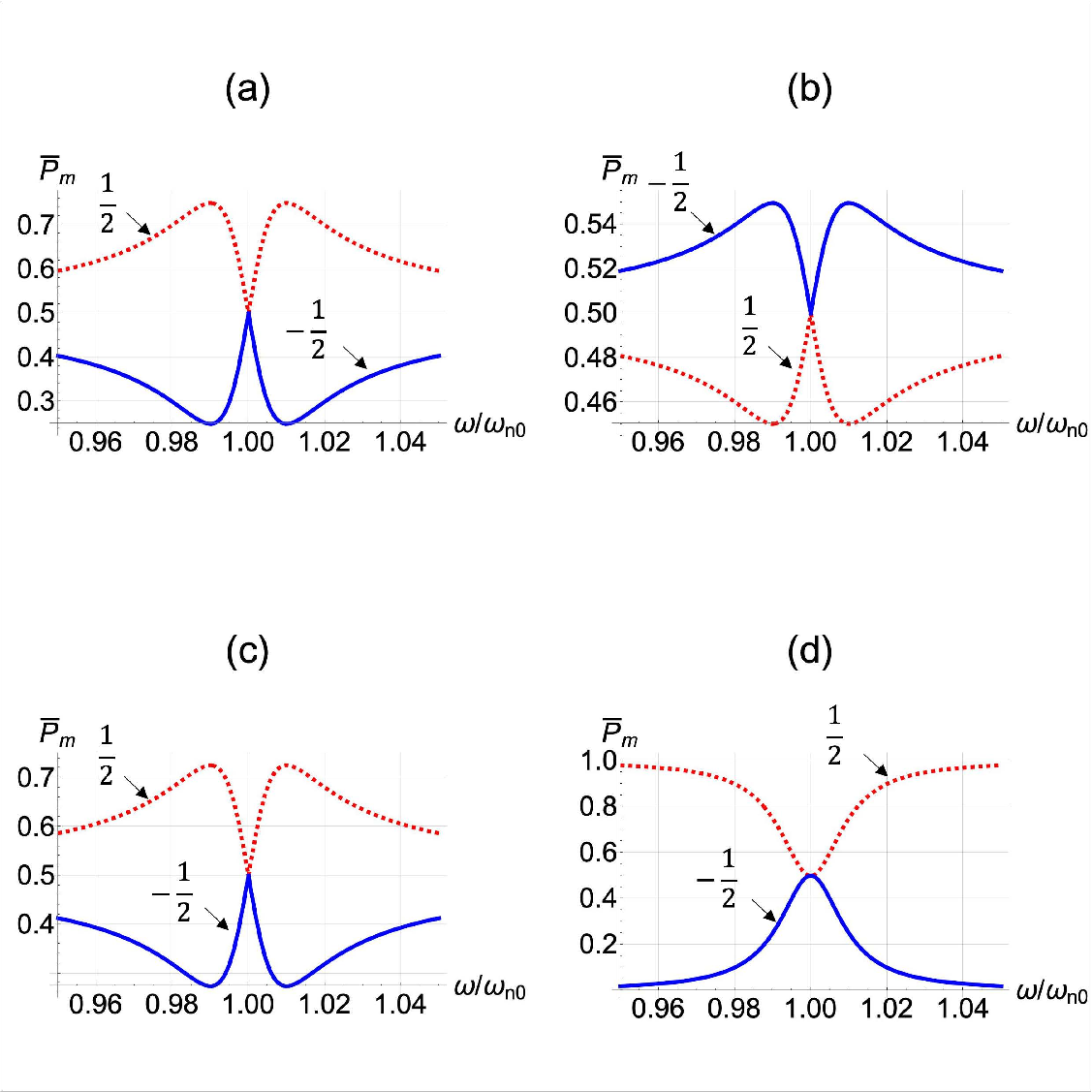}
\caption{Plots of $\overline{P}_m(\omega/\omega_{n0})$ for $I=\frac{1}{2}$ and $0.95\le\omega/\omega_{n0}\le1.05$.  $m=\frac{1}{2}$:  (dotted red), and $m=-\frac{1}{2}$: (solid blue).  This figure also applies for $J=\frac{1}{2}$ with $m$ and $\omega_{n0}$ respectively replaced by $\overline{m}$ and 
$\omega_{e0}$.}}
\end{figure}

\begin{figure}
\center{\includegraphics[width=0.49\textwidth]{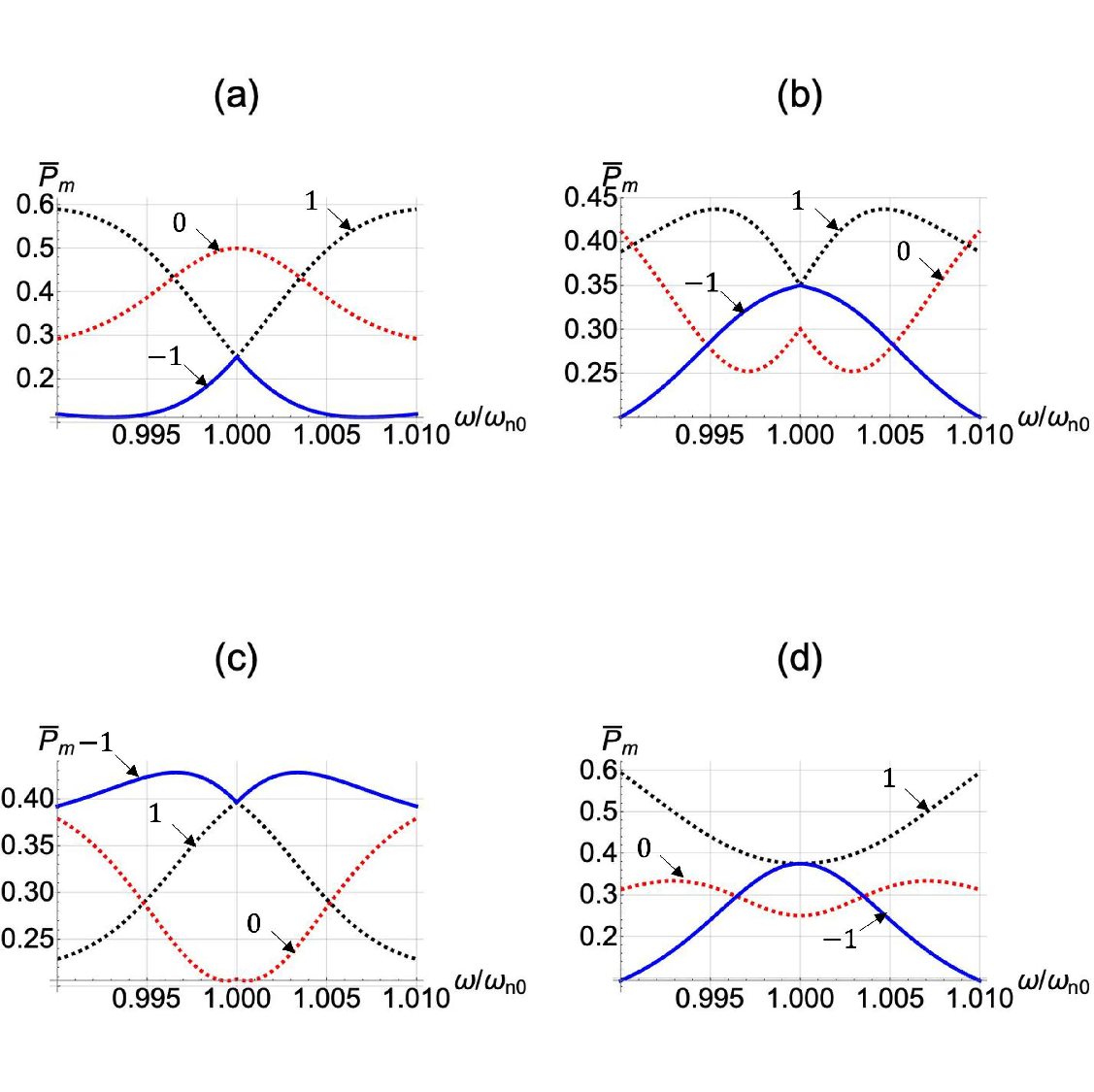}
\caption{Plots of $\overline{P}_m(\omega/\omega_{n0})$ for $I=1$ and $0.99\le\omega/\omega_{n0}\le1.01$. $m=1$: (dotted black), $m=0$: (dotted red), and $m=-1$: (solid blue).  This figure also applies for $J=1$ with $m$ and $\omega_{n0}$ respectively replaced by $\overline{m}$ and $\omega_{e0}$.}}
\end{figure}

\begin{figure}
\center{\includegraphics[width=0.49\textwidth]{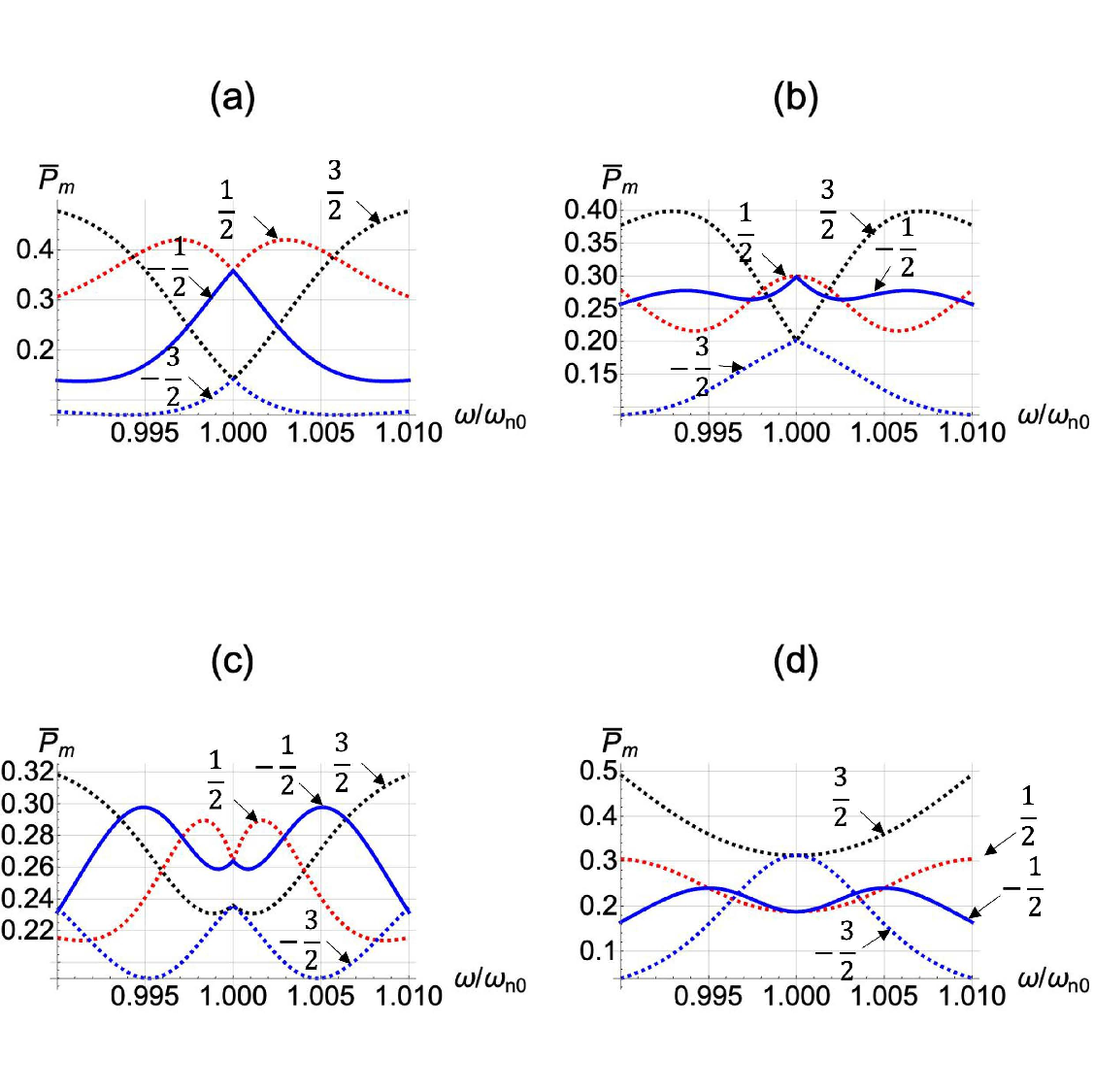}
\caption{Plots of $\overline{P}_m(\omega/\omega_{n0})$ for $I=\frac{3}{2}$ and $0.99\le\omega/\omega_{n0}\le1.01$.  $m=\frac{3}{2}$: (dotted black), $m=\frac{1}{2}$: (dotted red), $m=-\frac{1}{2}$: (solid blue), and $m=-\frac{3}{2}$: (dotted blue).  This figure also applies for $J=\frac{3}{2}$ with $m$ and 
$\omega_{n0}$ respectively replaced by $\overline{m}$ and $\omega_{e0}$.}}
\end{figure}

\begin{figure}
\center{\includegraphics[width=0.49\textwidth]{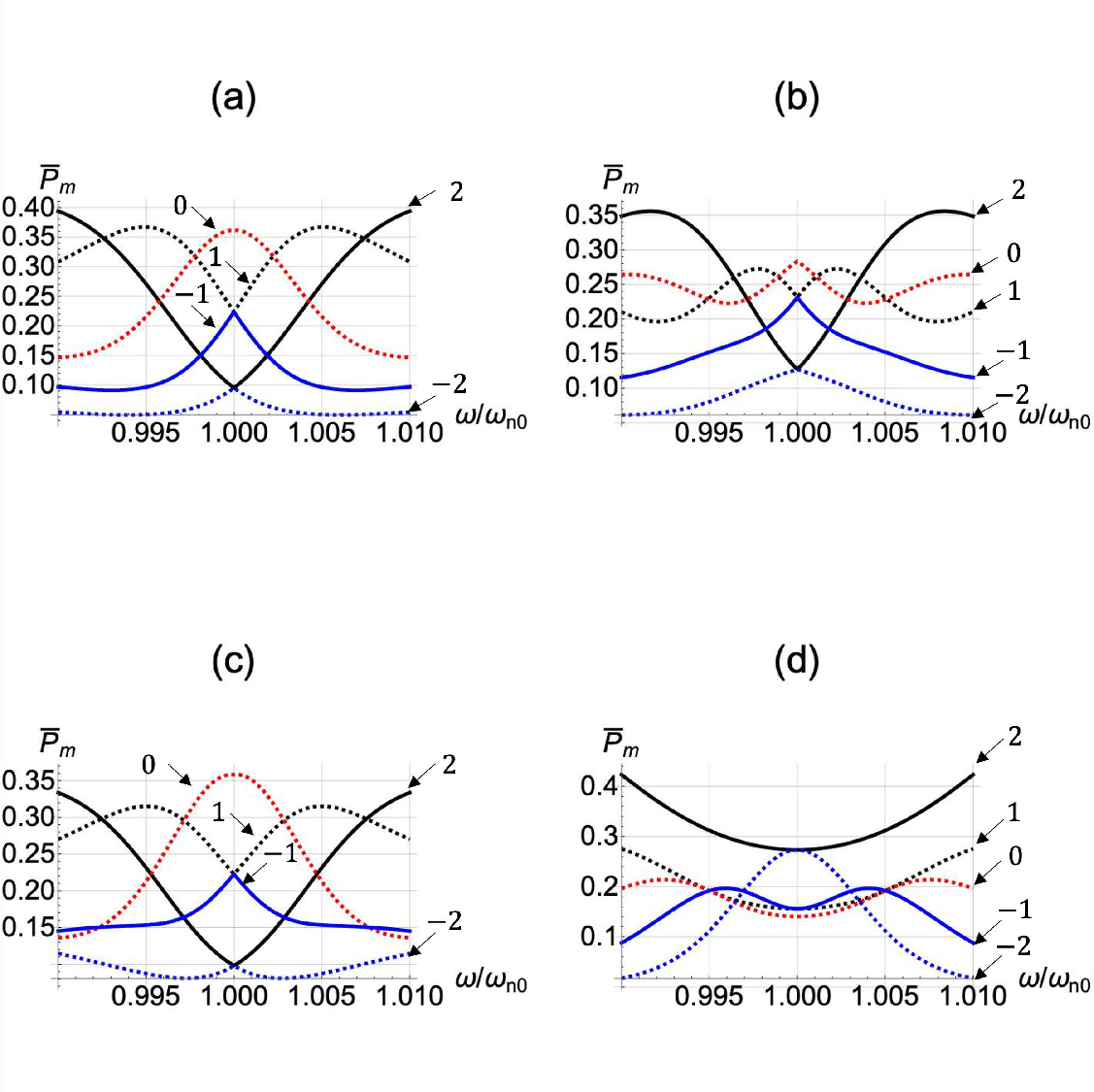}
\caption{Plots of $\overline{P}_m(\omega/\omega_{n0})$ for $I=2$ and $0.99\le\omega/\omega_{{\color{red}n}0}\le1.01$. $m=2$: (solid black), $m=1$: (dotted black), $m=0$: (dotted red), $m=-1$: (solid blue), and $m=-2$: (dotted blue).  This figure also applies for $J=2$ with $m$ and $\omega_{n0}$ respectively replaced by $\overline{m}$ and $\omega_{e0}$.
}}
\end{figure}

\section{Perturbations about the bare Hamiltonian}

For nuclei with $I\ge 1$, the nuclear electric quadrupole moment can exist \cite{Ramsey,RamseyII,Pyykkoe}, and in many cases, can dramatically broaden the NMR transitions. This is especially true for $^{14}$N, which has a 99.6\% natural abundance, but it has been particularly difficult to observe in molecules with standard NMR techniques until rather recently, when the sample was dissolved in  a variety of solvents \cite{Deng,Cavadini,Veinberg,Zeman}. However, $^{14}$N is an essential isotope in the constituent organic molecules adenine, cytosine, guanine, and thymine of deoxyribonuclei acid (DNA). Table I contains a list of the dipole and quadrupole moments and the nuclear spin and parity values for the elemental constituents of those molecules, as well as the chemically possible defect isotopes, along with some other isotopes relevant for  vertebrates, particularly for humans.  In the first-row atoms \cite{Trees,AHSI,AHSII,AHSIII,Armstrong}, the interaction of the electrons with the nucleus gives rise to hyperfine structure in the electronic atomic structure.  Hund's rules  for the electronic state hieroglyphic notation $^{2S+1}$L$_{J}$, \cite{Rohlf,Griffiths} along with the overall antisymmetry of the electronic states for an odd number of electrons, indicate that for only filled $n=1$ and partially filled $n=2$ atomic level states, the electronic ground state of atomic N is $^4$S$_{\frac{3}{2}}$, and  excited states successively higher in energy are  $^2$D$_{\frac{3}{2}}$,  $^2$D$_{\frac{5}{2}}$, $^2$P$_{\frac{1}{2}}$, $^2$P$_{\frac{3}{2}}$,  and $^2$S$_{\frac{1}{2}}$. Thus, there are 6 low-lying electronic states of N with $J=\frac{5}{2},\frac{3}{2}$, and $\frac{1}{2}$.

Thus, for atoms, one can treat the low-lying states accurately by assuming the electronic parts of their nuclear moments can be treated by their $J$ values, and rank-ordered in terms of their energies.  Of course in molecules such as adenine, cytosine, guanine, and thymine, with multiple C, N, O, and H atoms, one might question whether the overall molecular structure could have a quantized angular momentum.  We will return to this question in the following.

The nuclear electric quadrupole moment for $I\ge 1$ in atoms contains terms proportional to the three electric field gradients, $\frac{\partial {\bm E}}{\partial x_i}$, which are highly anisotropic.  This was treated perturbatively using the standard interaction representation starting with the time-independent magnetic field Hamiltonian $H_0=\mu_NI_z$ \cite{Urban}.  However, as Gottfried emphasized, such treatments cannot possibly work at or near to resonance \cite{Gottfried}.

For atoms, we assume that the nuclear total angular momentum states $|I,m\rangle$ with  $-I\le m\le I$ and operators ${\bm I}^2$ and $I_z$ satisfy Eq. (\ref{commutator}), with $-I\le m\le I$, and the analogous electronic states $|J,\overline{m}\rangle$ with $-J\le\overline{m}\le J$ satisfy
\begin{eqnarray}
{\bm J}^2|J,\overline{m}\rangle&=&\hbar^2J(J+1)|J,\overline{m}\rangle,\\
J_z|J,\overline{m}\rangle&=&\overline{m}\hbar|J,\overline{m}\rangle,
\end{eqnarray}
and the $J_i$ operators obey the Lie algebra, Eq. (\ref{commutator}), and are  independent of the nuclear $I_j$ operators,
 as has commonly been assumed in the literature \cite{Trees,AHSI,AHSII,AHSIII,Armstrong}.
In early atomic hyperfine structure calculations of the first row atoms, the electronic total angular momentum ${\bm J}$ is coupled to the nuclear spin ${\bm I}$ in  powers of ${\bm I}\cdot{\bm J}$ \cite{Trees,AHSI,AHSII,AHSIII,Armstrong}.  The nuclear-electronic magnetic dipole, electric  quadrupole, and  magnetic octupole interactions may be written as the scalar operators
\begin{eqnarray}
H_D&=&A(I,J){\bm I}\cdot{\bm J}/\hbar^2,\label{HD}\\
H_Q&=&B(I,J)\Bigl[({\bm I}\cdot{\bm J}/\hbar^2)^2+\frac{1}{2}({\bm I}\cdot{\bm J}/\hbar^2)\nonumber\\
& &-\frac{1}{3}I(I+1)J(J+1)\Bigr],\nonumber\\
& &\label{HQ}\\
H_O&=&C(I,J)\Bigl[({\bm I}\cdot{\bm J}/\hbar^2)^3+2({\bm I}\cdot{\bm J}/\hbar^2)^2\nonumber\\
& &+\frac{1}{5}({\bm I}\cdot{\bm J}/\hbar^2)[-3I(I+1)J(J+1)+I(I+1)\nonumber\\
& &+J(J+1)+3]-5I(I+1)J(J+1)\Bigr]\label{HO},
\end{eqnarray}
which  are respectively proportional to linear, quadratic, and cubic functions of $({\bm I}\cdot{\bm J})$, where $A(I,J)$, $B(I,J)$, and $C(I,J)$ are energies that are real functions of $I$ and $J$  respectively proportional to the nuclear magnetic dipole,  electric quadrupole, and magnetic octupole moments that separately contain the integrations over the nuclear and electronic structural variables in the $|I,I\rangle\times|J,J\rangle$ state that factor due to the Wigner-Eckart theorem \cite{Ramsey,RamseyII,BS,Armstrong}.  Similarly the electric hexadecapole moment can be written as a quartic function of $({\bm I}\cdot{\bm J})$ \cite{Ramsey,BS,Armstrong}.  Hence, in the absence of such moments, there is a separation of the electronic and the nuclear quantum variables, and at least for atoms, the couplings are proportional to powers of ${\bm I}\cdot{\bm J}$.  Thus for atoms, it has been standard to treat the overall electronic spin with the quantum operator ${\bm J}$, which is entirely analogous to the nuclear spin operator ${\bm I}$.

We note that the nuclear moment perturbations were studied in the absence of an applied magnetic field long ago to include their effects upon  the  atomic hyperfine structure of atoms \cite{Trees,AHSI,AHSII,AHSIII,Armstrong}.  In those cases,   the operators ${\bm I}$ and ${\bm J}$, as well as the combined ${\bm F}={\bm I}+{\bm J}$ gave good quantum numbers, and the nuclear moment perturbations could be written in terms of the single quantum number $K=F(F+1)$ \cite{Trees,AHSI,AHSII,AHSIII,Armstrong}.  However, with the three-component  applied magnetic fields, that symmetry is completely broken in all three directions, but the scalar operator ${\bm I}\cdot{\bm J}$ is well defined, provided that $I$ and $J$ are both good quantum numbers.

Thus, starting with the bare Hamiltonians for the nuclei and the electronic cloud of an atom, we assume the total Hamiltonian can be written as
\begin{eqnarray}
\hat{H}_{ne}(t)&=&\hat{H}_{n}(t)\otimes{\bm 1}_e+{\bm 1}_n\otimes\hat{H}_e(t)+\hat{\delta V}_{ne},\label{Htotal}\\
\hat{H}_n(t)&=&\hat{H}^0_{n}(t)+\hat{V}_n(t),\label{nuclearH}\\
\hat{H}_e(t)&=&\hat{H}^0_{e}(t)+\hat{V}_e(t),\label{electronH}\\
\hat{\delta V}_{ne}&=&\sum_{q=0}^{2I}K_q\left[{\bm I}\cdot{\bm J}/\hbar^2\right]^{q}{\bm 1}_n\otimes{\bm 1}_e,
\end{eqnarray}
where ${\bm 1}_n$ and ${\bm 1}_e$ are identity matrices of respective ranks $2I+1$ and $2J+1$, the $K_q$ are real functions of $I,J$, and the starting (or bare) Hamiltonian consists of $\hat{H}^0_{n}(t)$ given by Eq. (\ref{Hn}) and $\hat{H}^0_{e}(t)$ given by Eq. (\ref{He}).

We note that $|\psi_{ne}(t)\rangle$ contains components of the vector operators ${\bm I}$ and ${\bm J}$,  which can  respectively be represented by a matrices of rank $2I+1$ and $2J+1$ that operate on the nuclear and electronic angular momentum subspaces of those ranks.  The matrix representation of the combined  Hamiltonian $\hat{H}_{ne}(t)$ presented in the following is a $(2I+1)\times(2I+1)$ matrix representing the nuclear spin $I$ states multiplied by an independent  $(2J+1)\times(2J+1)$ matrix representing the electronic angular momentum $J$ states, and $|\psi_{ne}(t)\rangle$ is a rank $2I+2J+2$ column vector in Nambu representation, the upper $2I+1$ and lower $2J+1$ elements are acted upon by the respective independent rank $2I+1$ and $2J+1$ nuclear and electronic Hamiltonian matrix forms.

It is then easy to show that the above bare  $|\psi^{0}_{ne}(t)\rangle$ for the electrons and nuclei in atoms satisfies the Schr{\"o}dinger equation for the full  Hamiltonian,
\begin{eqnarray}
\hat{H}^{0}_{ne}(t)&=&\hat{H}^0_{n}(t)\otimes{\bm 1}_e\nonumber\\ & &+{\bm 1}_n\otimes\hat{H}^0_{e}(t),\label{HRWAneoft}\\
i\hbar\frac{\partial}{\partial t}|\psi^{0}_{ne}(t)\rangle&=&\hat{H}^{0}_{ne}(t)|\psi^{0}_{ne}(t)\rangle.
\end{eqnarray}

Thus, we therefore consider two types of perturbations $\hat{V}(t)$:  those that are variations in the explicit periodic time-dependence of the actual experimental applied magnetic field, and those that are time-independent natural nuclear moment perturbations that couple the nuclear spin operator ${\bm I}$ with the surrounding electronic spin  operator ${\bm J}$.

Depending upon the type of experiment, there will be a time-dependent perturbation of the difference between the experimental time-dependent field and the RWA time-dependent field.  These effect both the nuclei and the electrons. For a general time-periodic field form in the ${\bm x}$ direction, we have
\begin{eqnarray}
\hat{\delta V}_1(t)&=&\hat{\delta V}_{n1}(t)\otimes{\bm 1}_e+{\bm 1}_n\otimes\hat{\delta V}_{e1}(t),\label{deltaV1}\\
\hat{\delta V}_{n1}(t)&=&-\omega_{n1}I_y\sin(\omega t)+\omega_{n1}I_x\sum_{p=2}^{\infty}a_p\cos[p\omega(t-t_0)],\nonumber\\
& &\label{deltaVn1}\\
\hat{\delta V}_{e1}(t)&=&-\omega_{e1}J_y\sin(\omega t)+\omega_{e1}J_x\sum_{p=2}^{\infty}a_p\cos[p\omega(t-t_0)],\nonumber\\
& &\label{deltaVe1}
\end{eqnarray}
where any wave form can be calculated using standard Fourier series techniques \cite{Boas}.

For the magic-angle spinning case \cite{Hong}, we have
\begin{eqnarray}
\hat{\delta V}_2(t)&=&\hat{\delta V}_{n2}(t)\otimes{\bm 1}_e+{\bm 1}_n\otimes\hat{\delta V}_{e2}(t),\label{deltaV2}\\
\hat{\delta V}_{n2}(t)&=&2\omega_{n1}\sin\theta\cos(\omega t)(I_z\cos\theta-I_x\sin\theta),\label{deltaVn2}\nonumber\\
& &\\
\hat{\delta V}_{e2}(t)&=&2\omega_{e1}\sin\theta\cos(\omega t)(J_z\cos\theta-J_x\sin\theta).\label{deltaVe2}
\end{eqnarray}
For each of the two experiments $i=1,2$, the full perturbation that includes the nuclear moments and either of these time-dependent experiments may then be written as
\begin{eqnarray}
\hat{\delta V}_i(t)&=&\hat{\delta V}_{ni}(t)\otimes{\bm 1}_e+{\bm 1}_n\otimes\hat{\delta V}_{ei}(t)+\hat{\delta V}_{n,e}\label{deltaV},\\
\hat{\delta V}_{n,e}&=&
\sum_{q=0}^{2I}K_q({\bm I}\cdot{\bm J}/\hbar^2)^q{\bm 1}_n\otimes{\bm 1}_e,\label{nuclearmoments}
\end{eqnarray}
where $i=1,2$ for the two experiments and the $K_q$ are functions of the quantum numbers $I$ and $J$ and of the various constants for the nuclear moment and electronic  wave functions evaluated in the $|I,I\rangle$ and $|J,J\rangle$ states according to the Wigner-Eckart theorem \cite{Armstrong}.  All of the interactions between the nuclei and electrons in an atom, from the magnetic dipole, electric quadrupole, magnetic octupole, electric hexadecapole, etc., etc.,  are all fully represented by Eq. (\ref{nuclearmoments}).  We note that ${\bm I}\cdot{\bm J}$ is a scalar operator \cite{Trees,AHSI,AHSII,AHSIII,Armstrong}.

We are now ready to calculate the perturbations due to $\lambda\hat{\delta V}_i(t)$, each of which is assumed proportional to the dummy variable $\lambda$.  Letting
\begin{eqnarray}
|\psi_{ne}(t)\rangle&=&|\psi_{RWA,ne}(t)\rangle+\lambda|\psi_{1,ne}(t)\rangle\nonumber\\
& &+\lambda^2|\psi_{2,ne}(t)\rangle+\ldots,\\
\hat{H}_{ne}(t)&=&\hat{H}_{RWA,ne}(t)+\lambda\hat{\delta V}_i(t),
\end{eqnarray}
and requiring that the Schr{\"o}dinger equation be satisfied in each order $\ell$ of $\lambda$,
\begin{eqnarray}
i\hbar\frac{\partial}{\partial t}|\psi_{\ell,ne}(t)\rangle&=&\hat{H}_{ne}(t)|\psi_{\ell,ne}(t)\rangle,
\end{eqnarray}
we have to ${\cal O}(\lambda^0)$ that
\begin{eqnarray}
i\hbar\frac{\partial}{\partial t}|\psi^{0}_{ne}(t)\rangle&=&\hat{H}^{0}_{ne}(t)|\psi^{0}_{ne}(t)\rangle.\label{orderzero}
\end{eqnarray}
To ${\cal O}(\lambda)$, we have to satisfy
\begin{eqnarray}
i\hbar\frac{\partial}{\partial t}|\psi_{1,ne}(t)\rangle&=&\hat{H}^{0}_{ne}(t)|\psi_{1,ne}(t)\rangle+\hat{\delta V}_i(t)|\psi^{0}_{ne}(t)\rangle,\label{orderone}\nonumber\\
\end{eqnarray}
and to ${\cal O}(\lambda^{\ell})$ where $\ell\ge 1$ is an integer, the equation to satisfy is
\begin{eqnarray}
i\hbar\frac{\partial}{\partial t}|\psi_{\ell,ne}(t)\rangle&=&\hat{H}^{0}_{ne}(t)|\psi_{\ell,ne}(t)\rangle+\hat{\delta V}_i(t)|\psi_{\ell-1,ne}(t)\rangle.\label{orderell}\nonumber\\
\end{eqnarray}
We first transform the expression to ${\cal O}(\lambda^0)$, Eq. (\ref{orderzero}).
We write
\begin{eqnarray}
&& e^{i(\beta_nI_y{\bm 1}_e+\beta_e{\bm 1}_nJ_y)/\hbar}e^{i\omega t(I_z{\bm 1}_e+{\bm 1}_nJ_z)/\hbar}\Bigl(i\hbar\frac{\partial}{\partial t}\Bigr)\nonumber\\
& &\times e^{-i\omega t(I_z{\bm 1}_e+{\bm 1}_nJ_z)/\hbar}e^{-i(\beta_nI_y{\bm 1}_e+\beta_e{\bm 1}_nJ_y)/\hbar}\label{tildepartialt}\\
& &\times e^{i(\beta_nI_y{\bm 1}_e+\beta_e{\bm 1}_nJ_y)/\hbar}e^{i\omega t(I_z{\bm 1}_e+{\bm 1}_nJ_z)/\hbar}|\psi^{0}_{ne}(t)\rangle\nonumber\\
&=&e^{i(\beta_nI_y{\bm 1}_e+\beta_e{\bm 1}_nJ_y)/\hbar}e^{i\omega t(I_z{\bm 1}_e+{\bm 1}_nJ_z)/\hbar}\hat{H}^{0}_{ne}(t)\nonumber\\
& &\times e^{-i\omega t(I_z{\bm 1}_e+{\bm 1}_nJ_z)/\hbar}e^{-i(\beta_nI_y{\bm 1}_e+\beta_e{\bm 1}_nJ_y)/\hbar}\nonumber\\
& &\times e^{i(\beta_nI_y{\bm 1}_e+\beta_e{\bm 1}_nJ_y)/\hbar}e^{i\omega t(I_z{\bm 1}_e+{\bm 1}_nJ_z)/\hbar}|\psi^{0}_{ne}(t)\rangle,\label{transforms}\nonumber\\
\end{eqnarray}
where $\hat{H}^{0}_{ne}(t)$ is given by Eq. (\ref{barewavefunction}).  We note that Eq. (\ref{transforms}) could be rewritten as
\begin{eqnarray}
i\hbar\widetilde{\frac{\partial}{\partial t}}|\widetilde{\psi}^{0}_{ne}(t)\rangle&=&\widetilde{\hat{H}^{0}}_{ne}(t)|\widetilde{\psi}^{0}_{ne}(t)\rangle,
\end{eqnarray}
where
\begin{eqnarray}
\widetilde{\frac{\partial}{\partial t}}&=&e^{i(\beta_nI_y{\bm 1}_e+\beta_e{\bm 1}_nJ_y)/\hbar}e^{i\omega t(I_z{\bm 1}_e+{\bm 1}_nJ_z)/\hbar}\Bigl(\frac{\partial}{\partial t}\Bigr)\nonumber\\
& &\times e^{-i\omega t(I_z{\bm 1}_e+{\bm 1}_nJ_z)/\hbar}e^{-i(\beta_nI_y{\bm 1}_e+\beta_e{\bm 1}_nJ_y)/\hbar},\label{widetildepartialt}\nonumber\\
& &\\
\widetilde{\hat{H}^{0}}_{ne}(t)&=&e^{i(\beta_nI_y{\bm 1}_e+\beta_e{\bm 1}_nJ_y)/\hbar}e^{i\omega t(I_z{\bm 1}_e+{\bm 1}_nJ_z)/\hbar}\nonumber\\
& &\times\hat{H}^{0}_{ne}(t)\nonumber\\
& &\times e^{-i\omega t(I_z{\bm 1}_e+{\bm 1}_nJ_z)/\hbar}e^{-i(\beta_nI_y{\bm 1}_e+\beta_e{\bm 1}_nJ_y)/\hbar},\label{widetildeHRWAne}\nonumber\\
&&\\
|\widetilde{\psi}^{0}_{ne}(t)\rangle&=&e^{i(\beta_nI_y{\bm 1}_e+\beta_e{\bm 1}_nJ_y)/\hbar}e^{i\omega t(I_z{\bm 1}_e+{\bm 1}_nJ_z)/\hbar}\nonumber\\
& &\times|\psi^{0}_{ne}(t)\rangle,\label{widetildepsiRWAne}\nonumber\\
\end{eqnarray}
which is equivalent to the bare Schr{\"o}dinger equation,
\begin{eqnarray}
i\hbar\frac{\partial}{\partial t}|\widetilde{\psi}^{0}_{ne}(t)\rangle&=&\hat{H}_0|\widetilde{\psi}^{0}_{ne}(t)\rangle,\label{SWEne}
\end{eqnarray}
where the {\it time-independent} bare Hamiltonian $\hat{H}_0$ is given by
\begin{eqnarray}
\hat{H}_0&=&\Omega_nI_z\otimes{\bm 1}_e+\Omega_e{\bm 1}_n\otimes J_z.
\end{eqnarray}
From Eqs. (\ref{barewavefunction}) and (\ref{widetildepsiRWAne}), it is evident that
\begin{eqnarray}
|\widetilde{\psi}^{0}_{ne}(t)\rangle&=&e^{-it(\Omega_nI_z{\bm 1}_e+\Omega_n{\bm 1}_nJ_z)/\hbar}e^{i(\beta_nI_y{\bm 1}_e+\beta_e{\bm 1}_nJ_y)/\hbar}\nonumber\\
& &\times|\psi^{0}_{ne}(0)\rangle\\
&=&e^{-i\hat{H}_0t/\hbar}|\psi^{0}_{ne,r}(0)\rangle_S,\label{RWASWE}
\end{eqnarray}
where the doubly-rotated time-independent RWF wave function in the Schr{\"o}dinger picture is
\begin{eqnarray}
|\psi^{0}_{ne,r}(0)\rangle_S&=&e^{i(\beta_nI_y{\bm 1}_e+\beta_e{\bm 1}_nJ_y)/\hbar}|\psi^{0}_{ne}(0)\rangle.\nonumber\\
\label{doublyrotatedpsi}
\end{eqnarray}

Since the original laboratory frame basis given by Eq. (\ref{psineinitial}) leads to a rather complicated expression for the ground state energy, it is useful to make the change of basis to the doubly-rotated frame in the Schr{\"o}dinger picture, letting
\begin{eqnarray}
|\psi^{0}_{ne,r}(0)\rangle_S&\equiv&\sum_{m=-I}^I\sum_{\overline{m}=-J}^J\tilde{C}_m(0)\tilde{D}_{\overline{m}}(0)\nonumber\\
& &\times|I,m\rangle\otimes|J,\overline{m}\rangle,\label{rotatedbasis}
\end{eqnarray}
where for normalization, we require
\begin{eqnarray}
\sum_{m=-I}^I|\tilde{C}_m(0)|^2&=&\sum_{\overline{m}=-J}^J|\tilde{D}_{\overline{m}}(0)|^2=1.
\end{eqnarray}
The coefficients $\tilde{C}_m(0)$ and $\tilde{D}_{\overline{m}}(0)$ in the doubly-rotated new basis are easily seen to be given by
\begin{eqnarray}
\tilde{C}_m(0)&=&\sum_{m'=-I}^I\langle I,m|e^{i\beta_nI_y/\hbar}|I,m'\rangle C_{m'}(0)\nonumber\\
&=&\sum_{m'=-I}^Id_{m,m'}^{(I)*}(\beta_n)C_{m'}(0),\label{Cm0}\\
\tilde{D}_{\overline{m}}(0)&=&\sum_{\overline{m}'=-J}^J\langle J,\overline{m}|e^{i\beta_eJ_y/\hbar}|J,\overline{m}'\rangle D_{\overline{m}'}(0)\nonumber\\
&=&\sum_{\overline{m}'=-J}^Jd_{\overline{m},\overline{m}'}^{(J)*}(\beta_e)D_{\overline{m}'}(0),\label{Dm0}
\end{eqnarray}
where  $d_{m',m}^{(I)}(\beta_n)$ is given in textbooks \cite{SN3,Griffiths,GY} and explicitly presented in Eq. (\ref{dsm'm}).  Of course, $d_{\overline{m},\overline{m}'}^{(J)*}(\beta_e)$ is readily obtained from  $d_{m',m}^{(I)}(\beta_n)$ by the appropriate substitutions and complex conjugation.

Then, in this doubly-rotated basis, the energy of a single ``non-interacting'' state, which only interacts with the applied magnetic field, is
\begin{eqnarray}
\tilde{E}^{(0)}_{ne}(m,\overline{m})&=&\hbar(m\Omega_n+\overline{m}\Omega_e),\label{E0}
\end{eqnarray}
and the most general form of the non-interacting  state energy is
\begin{eqnarray}
\tilde{E}^{(0)}_{ne}&=&\hbar\Bigl(\Omega_n\sum_{m=-I}^Im|\tilde{C}_{m}(0)|^2+\Omega_e\sum_{\overline{m}=-J}^J\overline{m}|\tilde{D}_{\overline{m}}(0)|^2\Bigr).\label{groundstateenergy}\nonumber\\
\end{eqnarray}

With this change of basis, we are now in the position to  use standard  perturbation theory \cite{SN3,Gottfried,GY,Griffiths}.  Transforming Eq. (\ref{orderell}) as for ${\cal O}(\lambda^{0})$, we obtain
\begin{eqnarray}
i\hbar\widetilde{\frac{\partial}{\partial t}}|\widetilde{\psi}_{\ell,ne}(t)\rangle&=&\widetilde{\hat{H}}_{RWA,ne}(t)|\widetilde{\psi}_{\ell,ne}(t)\rangle+\widetilde{\hat{\delta V}_i(t)}|\widetilde{\psi}_{\ell-1,ne}(t)\rangle,\nonumber\\
\label{transformedperturbations}
\end{eqnarray}
where
\begin{eqnarray}
|\widetilde{\psi}_{\ell,ne}(t)\rangle&=&e^{-it(\Omega_nI_z{\bm 1}_e+\Omega_n{\bm 1}_nJ_z)/\hbar}e^{i(\beta_nI_y{\bm 1}_e+\beta_e{\bm 1}_nJ_y)/\hbar}\nonumber\\
& &\times|\psi_{\ell,ne}(0)\rangle,\\
\widetilde{\hat{\delta V}_i(t)}&=&e^{i(\beta_nI_y{\bm 1}_e+\beta_e{\bm 1}_nJ_y)/\hbar}e^{i\omega t(I_z{\bm 1}_e+{\bm 1}_nJ_z)/\hbar}\hat{\delta{V}}_i(t)\nonumber\\
& &\times e^{-i\omega t(I_z{\bm 1}_e+{\bm 1}_nJ_z)/\hbar}e^{-i(\beta_nI_y{\bm 1}_e+\beta_e{\bm 1}_nJ_y)/\hbar}.\nonumber\\
&&
\end{eqnarray}
Expressions for the $\widetilde{\hat{\delta V}_i(t)}$ are given in the Appendix.

\subsection{Nuclear moment perturbations in atoms}

Recently, there has been an increased interest in the measurement of atomic nuclear moments, with most of the work being on measurements of the electric quadrupole moment and the magnetic octupole moment \cite{Allegrini,IOC,Budker,Stejskal,Frydman,Gerginov,BeloyI,BeloyII,Moudrakovski,Singh,Ashbrook,Leroy,Xiao,deGrooteI,Smith,deGrooteII,Li,Rahaman,Persson}  Hence a new experimental method that could lead to improved accuracy of such moment measurements is presently warranted.

It is easily shown that the untransformed nuclear moment interactions may be written as in Eq. (\ref{Delta}),
\begin{eqnarray}
\hat{\delta V}_{ne}&=&\sum_{q=0}^{2I}K_q({\bm I}\cdot{\bm J}/\hbar^2)^q,\label{deltaVne}
\end{eqnarray}
where
\begin{eqnarray}
{\bm I}\cdot{\bm J}&=&I_zJ_z+\frac{1}{2}(I_{-}J_{+}+I_{+}J_{-}),
\label{IdotJ}
\end{eqnarray}
and where $J_{\pm}=J_x\pm iJ_y$ and $I_{\pm}=I_x\pm iI_y$. Provided that $I$ and $J$ are good quantum numbers, such as for atoms in the absence of the nuclear moment interactions,  the entire infinite set of these transformed nuclear moment interactions are independent of $t$, and can in principle be solved to arbitrary order in time-independent perturbation theory.
Then the transformed nuclear moment interactions may be written as
\begin{eqnarray}
\widetilde{\hat{\delta V}_{ne}}&=&e^{i(\beta_nI_y{\bm 1}_e+\beta_e{\bm 1}_nJ_y)/\hbar}e^{i\omega t(I_z{\bm 1}_e+{\bm 1}_nJ_z)/\hbar}\hat{\delta{V}}_{ne}\nonumber\\
& &\times e^{-i\omega t(I_z{\bm 1}_e+{\bm 1}_nJ_z)/\hbar}e^{-i(\beta_nI_y{\bm 1}_e+\beta_e{\bm 1}_nJ_y)/\hbar}.\>\>\>\>\>\>\>
\end{eqnarray}
Now we treat this perturbation as follows:  we first transform $\hat{\delta V}_{ne}$ with respect to the two rotations about the $z$ axis by the same angle $-\omega t$.  It is easy to show that
\begin{eqnarray}
e^{i\omega t(I_z{\bm 1}_e+{\bm 1}_nJ_z)/\hbar}({\bm I}\cdot{\bm J})e^{-i\omega t(I_z{\bm 1}_e+{\bm 1}_nJ_z)/\hbar}&=&({\bm I}\cdot{\bm J}),\nonumber\\
\end{eqnarray}
which is independent of $t$.   This can be done by generalizing the procedures from Eq. (\ref{Ixyofphi}) to Eq. (\ref{Ixyofphianswer}) for the rotated components of ${\bm I}$ to the analogous ones for the components of ${\bm J}$. It is also obviously true that when ${\bm I}$ and ${\bm J}$ are rotating about the same axis by the same angle, their scalar product is invariant under the rotation. Then one can similarly transform $({\bm I}\cdot{\bm J})^2$ by inserting
\begin{eqnarray}
{\bm 1}_n\otimes{\bm 1}_e&=&e^{i\omega t(I_z{\bm 1}_e+{\bm 1}_nJ_z)/\hbar}e^{-i\omega t(I_z{\bm 1}_e+{\bm 1}_nJ_z)/\hbar}
\end{eqnarray}
between the two factors of $({\bm I}\cdot{\bm J})$. Thus, all of the transformed nuclear moment interactions  are completely independent of the time. Then, the two rotation operator transformations can be applied to the wave functions in the  laboratory frame by changing the basis to the doubly-rotated frame.

For the leading non-trivial  examples of the measurements of the magnetic dipole, electric quadrupole, and magnetic octupole moments, the first order energy $\tilde{E}^{(1)}_{ne}$ is easiest to evaluate in the doubly-rotated frame, Eq. (\ref{rotatedbasis}), for which the wave function in the doubly-rotated frame is given by
\begin{eqnarray}
|\widetilde{\psi}_{0,ne,r}(0)\rangle&=&|\psi_{RWA,ne,r}\rangle_S.
\end{eqnarray}
The first-order perturbation for the nuclear moment interactions may therefore be written as
\begin{eqnarray}
\tilde{E}^{(1)}_{ne}&=&_{S}\langle\psi_{0,ne,r}(0)|\hat{\delta V}_{ne}|\psi_{0,ne,r}(0)\rangle_S,
\label{firstorder}
\end{eqnarray}
where $\hat{\delta V}_{ne}$ is given by Eq. (\ref{deltaVne}).  Note that the operations of $I_z,J_z, I_{\pm}$, and $J_{\pm}$ present in Eq. (\ref{IdotJ}) are given by \cite{Gottfried,SN3,GY}
\begin{eqnarray}
I_z|I,m\rangle&=&m\hbar|I,m\rangle,\\
J_z|J,\overline{m}\rangle&=&\overline{m}\hbar|J,\overline{m}\rangle\\
I_{\pm}|I,m\rangle&=&\hbar\sqrt{(I\mp m)(I\pm m+1)}|I,m\pm 1\rangle,\nonumber\\
&\equiv&\hbar\alpha^{I}_{m,\pm}|I,m\pm 1\rangle,\label{alphaImpm}\\
J_{\pm}|J,\overline{m}\rangle&=&\hbar\sqrt{(J\mp\overline{m})(J\pm\overline{m}+1)}|J,\overline{m}\pm 1\rangle\nonumber\\
&\equiv&\hbar\alpha^{J}_{\overline{m},\pm}|J,\overline{m}\pm 1\rangle\label{alphaJmpm}.
\end{eqnarray}

To evaluate the perturbation to the energy arising from  the magnetic dipole,  electric quadrupole, and magnetic octupole moments, we use the doubly-rotated reference frame, so that the bare Hamiltonian $\hat{H}_0$ is diagonal and independent of $t$.  Then, the $q=0, 1, 2, 3$ terms in $\hat{\delta V}_{ne}$ can be represented by the rank $2I+2J+2$ matrix $\hat{M}$ with elements
\begin{eqnarray}
\Bigl(\hat{M}\Bigr)^{\overline{m}',\overline{m}}_{m',m}&=&K_0\delta_{m',m}\delta_{\overline{m}',\overline{m}}+
K_1\Bigl[\delta_{m',m}\delta_{\overline{m}',\overline{m}}m\overline{m}\nonumber\\
& &+\frac{1}{2}\Bigl(\delta_{m',m-1}\delta_{\overline{m}',\overline{m}+1}\alpha^{I}_{m,-}\alpha^{J}_{\overline{m},+}\nonumber\\
& &
+\delta_{m',m+1}\delta_{\overline{m}',\overline{m}-1}\alpha^{I}_{m,+}\alpha^J_{\overline{m},-}\Bigr)\Bigr]\label{K1}\nonumber\\
& &+K_2\Biggl\{\delta_{m',m}\delta_{\overline{m}',\overline{m}}\biggl[m\overline{m}\Bigl(m\overline{m}-\frac{1}{2}\Bigr)\nonumber\\
& &\hskip30pt+\frac{1}{2}[I(I+1)-m^2][J(J+1)-\overline{m}^2]\biggr)\biggr]\nonumber\\
& &+\frac{1}{2}\Bigl(\delta_{m',m-1}\delta_{\overline{m}',\overline{m}+1}\alpha^I_{m,-}\alpha^J_{\overline{m},+}[2m\overline{m}+m-\overline{m}-1]\nonumber\\
& &+\delta_{m',m+1}\delta_{\overline{m}',\overline{m}-1}\alpha^{I}_{m,+}\alpha^J_{\overline{m},-}[2m\overline{m}+\overline{m}-m-1]\Bigr)\nonumber\\
& &+\frac{1}{4}\Bigl(\delta_{m',m-2}\delta_{\overline{m}',\overline{m}+2}\alpha^I_{m,-}\alpha^{I}_{m-1,-}\alpha^{J}_{\overline{m},+}\alpha^{J}_{\overline{m}+1,+}\nonumber\\
& &+\delta_{m',m+2}\delta_{\overline{m}',\overline{m}-2}\alpha^I_{m,+}\alpha^I_{m+1,+}\alpha^J_{\overline{m},-}\alpha^J_{\overline{m}-1,-}\Bigr)\Biggr\}\nonumber\\
& & +K_3\Biggl\{\delta_{m',m}\delta_{\overline{m}',\overline{m}}\Bigl(m^3\overline{m}^3\nonumber\\
& &+\frac{1}{4}(I-m)(I+m+1)(J+\overline{m})(J-\overline{m}+1)\times\nonumber\\
& &\times[2m\overline{m}+(m+1)(\overline{m}-1)]\nonumber\\
& &+\frac{1}{4}(I+m)(I-m+1)(J-\overline{m})(J+\overline{m}+1)\times\nonumber\\
& &\times[2m\overline{m}+(m-1)(\overline{m}+1)]\Bigr)\nonumber\\
& &+\frac{1}{8}\Bigl\{\delta_{m',m-1}\delta_{\overline{m}',\overline{m}+1}\alpha^{I}_{m,\-}\alpha^{J}_{\overline{m},+}\Bigl[4m^2\overline{m}^2\nonumber\\
& &+4(m-1)^2(\overline{m}+1)^2{m}(m-1)(\overline{m}+1)\nonumber\\
& &+(I-m)(I+m+1)(J+\overline{m})(J-\overline{m}+1)\nonumber\\
& &+(I+m)(I-m+1)(J-\overline{m})(J+\overline{m}+1)\nonumber\\
& &+(I+m-1)(I-m+2)(J+\overline{m}+2)(J-\overline{m}-1)\Bigr]\nonumber\\
& &+\delta_{m',m+1}\delta_{\overline{m}',\overline{m}-1}\alpha^{I}_{m,+}\alpha^{J}_{\overline{m},-}\Bigl[4m^2\overline{m}^2\nonumber\\
& &+4(m+1)^2(\overline{m}-1)^2+4m\overline{m}(m-1)(\overline{m}+1)\nonumber\\
& &+(I+m)(I-m+1)(J-\overline{m})(J+\overline{m}+1)\nonumber\\
& &+(I-m)(I+m+1)(J+\overline{m})(J-\overline{m}+1)\nonumber\\
& &+(I+m+2)(I-m-1)(J+\overline{m}-1)(J-\overline{m}+2)\Bigr]\nonumber\\
& &+\frac{1}{4}\delta_{m',m+2}\delta_{\overline{m}',\overline{m}-2}\alpha^{I}_{m,+}\alpha^{I}_{m+1,+}\alpha^{J}_{\overline{m},-}\alpha^{J}_{\overline{m}-1,-}\times\nonumber\\
& &\times[(m+2)(\overline{m}-2)+m\overline{m}+(m+1)(\overline{m}-1)]\nonumber\\
& &+\frac{1}{4}\delta_{m',m-2}\delta_{\overline{m}',\overline{m}+2}\alpha^{I}_{m,-}\alpha^{I}_{m-1,-}\alpha^{J}_{\overline{m},+}\alpha^{J}_{\overline{m}+1,+}\times\nonumber\\
& &\times[(m-2)(\overline{m}+2)+m\overline{m}+(m-1)(\overline{m}+1)]\nonumber\\
& &+\frac{1}{8}\delta_{m',m+3}\delta_{\overline{m}',\overline{m}-3}\alpha^{I}_{m,+}\alpha^{I}_{m+1,+}\alpha^{I}_{m+2,+}\times\nonumber\\
& &\times\alpha^{J}_{\overline{m},-}\alpha^{J}_{\overline{m}-1,-}\alpha^{J}_{\overline{m}-2,-}\nonumber\\
& &+\frac{1}{8}\delta_{m',m-3}\delta_{\overline{m}',\overline{m}+3}\alpha^{I}_{m,-}\alpha^{I}_{m-1,-}\alpha^{I}_{m-2,-}\times\nonumber\\
& &\times\alpha^{J}_{\overline{m},+}\alpha^{J}_{\overline{m}+1,+}\alpha^{J}_{\overline{m}+2,+}\Biggr\},
\end{eqnarray}
where in the simplest approximation in which the matrix elements do not depend upon $n$ and $L$, 
\begin{eqnarray}
K_0&=&-\frac{1}{3}B(I,J)I(I+1)J(J+1)\nonumber\\
& & -5C(I,J)I(I+1)J(J+1),\\
K_1&=&A(I,J)+\frac{1}{2}B(I,J)+\frac{1}{5}C(I,J),\\
K_2&=& B(I,J)+2C(I,J)\Bigl[-3I(I+1)J(J+1)\nonumber\\
& &+I(I+1)+J(J+1)+3\Bigr],\\
K_3&=&C(I,J),
\end{eqnarray}
and where $A(I,J)$, $B(I,J)$, and $C(I,J)$ are given elsewhere \cite{Armstrong}.  Then, the equation for the energy eigenvalues $E$ for the magnetic dipole, electric quadrupole, and magnetic octupole interactions for arbitrary $I,J$ is
\begin{eqnarray}
\det(\hat{H}_0+{\hat{M}-E{\bm 1}_n\otimes{\bm 1}_e})&=&0.
\end{eqnarray}

We note that some atoms can have couplings between different $J$ values, and the $K_i$ can often depend upon the other atomic quantum numbers such as $n$ and $L$.\cite{Allegrini}. We have not considered such cases, but the $K_i$ can be considered to have more general forms. 

The simplest example that includes the electric quadrupole moment but not the magnetic octupole moment is for $I=1$ and $J=1/2$, as for the ground $^2$S$_{1/2}$ state of $^2$H and the $^2$S$_{1/2}$ and $^2$P$_{1/2}$  excited states of $^{14}$N \cite{AHSIII}.  The rank-6 matrix $\hat{H}_0+\hat{M}$ has  horizontal elements $(m,\overline{m})$ and vertical elements $(m',\overline{m}')$.  One may define the upper left corner element to be $(m',\overline{m}')=(1,\frac{1}{2})$ and $(m,\overline{m})=(1,\frac{1}{2})$, and from far left to far right, the $(m,\overline{m})$ elements to be $(1,\frac{1}{2}), (1,-\frac{1}{2}), (0,\frac{1}{2}), (0,-\frac{1}{2}), (-1,\frac{1}{2})$, and $(-1,-\frac{1}{2})$.  The same order applies from top to bottom for the $(m',\overline{m}')$ states.  Then the six diagonal elements of $\hat{H}_0+\hat{M}$ from upper left corner to lower right corner are respectively $\hbar(\Omega_e/2+\Omega_n)+K_0+\frac{1}{2}K_1+\frac{1}{4}K_2$, $-\hbar(\Omega_e/2-\Omega_n)+K_0-\frac{1}{2}K_1+\frac{3}{4}K_2$, $\hbar\Omega_e/2+K_0+\frac{1}{2}K_2$, $-\hbar\Omega_e/2+K_0+\frac{1}{2}K_2$, $\hbar(\Omega_e/2-\Omega_n)+K_0-\frac{1}{2}K_1+\frac{3}{4}K_2$, and $\hbar(\Omega_e/2-\Omega_n)+K_0+\frac{1}{2}K_1+\frac{1}{4}K_2$.  There are four non-vanishing off-diagonal elements.  The second row $(m',\overline{m}')=(1,-\frac{1}{2})$ and third column $(m,\overline{m})=(0,\frac{1}{2})$ element is
\begin{eqnarray}
\Gamma&\equiv&\frac{1}{\sqrt{2}}\Bigl(K_1-\frac{K_2}{2}\Bigr).
\end{eqnarray}
 The third row $(m',\overline{m}')=(0,\frac{1}{2})$ and second column element $(m,\overline{m})=(1,-\frac{1}{2})$ is also $\Gamma$. The fourth row $(m',\overline{m}')=(0,-\frac{1}{2})$ and fifth column $(m,\overline{m})=(1-,\frac{1}{2})$ element is also $\Gamma$, and  the fifth row $(m',\overline{m}')=(1,\frac{1}{2})$ and fourth column $(m,\overline{m})=(0,-\frac{1}{2})$ element is  also $\Gamma$.

  Then, it is relatively easy to see that the determinant leads to two energy solutions linear in the $K_i$ and the four  solutions of two quadratic equations for $E$.  The two linear solutions for the energy  in the nuclear moments are
\begin{eqnarray}
E_1&=&\hbar(\Omega_e/2+\Omega_n)+K_0+K_1/2+K_2/4,\label{E1}\\
E_6&=&-\hbar(\Omega_e/2+\Omega_n)+K_0+K_1/2+K_2/4,\label{E2}
\end{eqnarray}
which correspond to the energies of the $(m,\overline{m})=(1,\frac{1}{2})$ and $(-1,-\frac{1}{2})$ states, respectively.
The four solutions of the two quadratic equations are
\begin{eqnarray}
E_{2,4}&=&K_0+\frac{1}{2}\biggl(\hbar\Omega_n+K_2-\Gamma/\sqrt{2}\nonumber\\
& &\pm\sqrt{(\hbar\Omega_e-\hbar\Omega_n+\Gamma/\sqrt{2})^2+4\Gamma^2}\biggr),\\
E_{3,5}&=&K_0+\frac{1}{2}\biggl(-\hbar\Omega_n+K_2-\Gamma/\sqrt{2}\nonumber\\
& &\pm\sqrt{(\hbar\Omega_e-\hbar\Omega_n-\Gamma/\sqrt{2})^2+4\Gamma^2}\biggr),
\end{eqnarray}
so that all of the $E_i$ are real.
$E_{2,4}$   correspond respectively for small $\Gamma$  to the  $(m,\overline{m})=(1,-\frac{1}{2})$ and $(0,\frac{1}{2})$ states, and $E_{3,5}$  correspond respectively for small $\Gamma$ to the  $(m,\overline{m})=(0,-\frac{1}{2})$ and $(-1,\frac{1}{2})$ states.

 For the $^4$S$_{\frac{3}{2}}$ ground state and $^2$D$_{\frac{3}{2}}$ excited state of $^{14}$N \cite{AHSIII}, the  $K_2$ contributions are generally stronger than for the $J=1/2$ states solved above.  The diagonal elements of $\hat{M}$ for the two $(m,\overline{m})=(\pm1,\mp\frac{3}{2})$ states are  $K_0-\frac{3}{2}K_1+\frac{15}{4}K_2$, those for the $(\pm1,\mp\frac{1}{2})$ states are $K_0-\frac{1}{2}K_1+\frac{9}{4}K_2$, those of the $(\pm1,\pm\frac{1}{2})$ states are $K_0 +\frac{1}{2}K_1+\frac{7}{4}K_2$, those for the $(\pm1,\pm\frac{3}{2})$ states are $K_0+\frac{3}{2}K_1+\frac{9}{4}K_2$, those for the $(0,\pm\frac{1}{2})$ states are $K_0+\frac{7}{4}K_2$, and those for the $(0,\pm\frac{3}{2})$ states are $K_0+\frac{3}{2}K_2$.

 Of course, there are also considerably more nonvanishing off-diagonal elements of $\hat{M}$ that contain $K_2$ than for the $^2$S$_{\frac{1}{2}}$ and  $^2$P$_{\frac{1}{2}}$ states. The $^2$D$_{\frac{5}{2}}$ excited state of $^{14}$N is even more complicated.  For this electronic state, the diagonal elements for the two $(m,\overline{m})=(\pm 1,\pm\frac{5}{2})$ states are $K_0+\frac{5}{2}K_1+\frac{25}{4}K_2$, those of the $(\pm 1,\pm\frac{3}{2})$ states are $K_0+\frac{3}{2}K_1+\frac{19}{4}K_2$, those for the $(\pm 1,\pm\frac{1}{2})$ states are $K_0+\frac{1}{2}K_1+\frac{17}{4}K_2$, those for the $(\pm 1,\mp\frac{1}{2})$ states are $K_0-\frac{1}{2}K_1+\frac{19}{4}K_2$, those for the $(\pm 1,\mp\frac{3}{2})$ states are $K_0-\frac{3}{2}K_1+\frac{25}{4}K_2$, those for the $(\pm 1,\mp\frac{5}{2})$ states are
$K_0-\frac{5}{2}K_1+\frac{35}{4}K_2$, those for the $(0,\pm\frac{5}{2})$ states are $K_0+\frac{5}{2}K_2$, those for the $(0,\pm\frac{3}{2})$ states are $K_0+\frac{13}{2}K_2$, and those for the $(0,\pm\frac{1}{2})$ states are $K_0+\frac{17}{2}K_2$.  There are also many different off-diagonal matrix elements. The full solutions of the electric quadrupole moment for each of the states of $^{14}$N  with filled $1s$ and $2s$ electronic orbitals to first order in perturbation theory will be presented elsewhere.

 Details of these first order results for specific atoms and additional higher order results for the electric quadrupole, magnetic octupole, and electric hexadecapole moment using this RWF magnetic field Hamiltonian as the actual experimental probe will be published elsewhere \cite{Stone}.  More complicated time-dependent interactions involving $\widetilde{\hat{\delta V}}_{n,i}(t)+\widetilde{\hat{\delta V}}_{e,i}(t)$ cases can be treated in the interaction representation once the transformation to the doubly-rotated basis is made \cite{Gottfried,Griffiths,SN3,GY}.  Specific examples of all of these cases will be presented in future publications.

An atom of particular interest in $^7$Li, which has $I=\frac{3}{2}$ and its ground electronic state is $^2$S$_{1/2}$. The two electronic excited states are $^2$P$_{1/2}$ and $^2$P$_{3/2}$.  From Table I, there have been nine measurements of its electric quadrupole moment, and they do not all agree with one another.  We suggest that the RWF magnetic field measurements could be useful to obtain more reliable results for the electric quadrupole and magnetic octupole moments for that atom.  A computer program for those computations is available \cite{Liudissertation}.

We remark that it is possible to also use the analogous extension of Gottfried's wave function $|\psi_{G,n}(t)\rangle$, given by Eq. (\ref{Gottfriedwavefunction}), to include the same perturbation expansion that includes the nuclear moments.  We could write the analogous electronic wave function as
\begin{eqnarray}
|\psi_{G,e}(t)\rangle&=&e^{-i\omega tJ_z/\hbar}e^{-i\hat{\bm n}_e\cdot{\bm\Omega}_e t/\hbar}|\psi_{G,e}(0)\rangle,\\
|\psi_{G,ne}(t)\rangle&=&|\psi_{G,n}(t)\rangle\otimes|\psi_{G,e}(t)\rangle\nonumber\\
&=&e^{-i\omega t(I_z{\bm 1}_e+{\bm 1}_nJ_z)/\hbar}\nonumber\\
& &\times e^{-it(\hat{\bm n}_n\cdot{\bm I}\Omega_n{\bm 1}_e+\Omega_e{\bm 1}_n\hat{\bm n}_e\cdot{\bm J})/\hbar}\nonumber\\
& &\times|\psi_{G,ne}(0)\rangle,\nonumber\\
\end{eqnarray}
where
\begin{eqnarray}
|\psi_{G,ne}(0)\rangle&=&|\psi_{G,n}(0)\rangle\otimes|\psi_{G,e}(0)\rangle,\\
\hat{\bm n}_e&=&\hat{\bm z}\cos(\beta_e)+\hat{\bm x}\sin(\beta_e),
\end{eqnarray}
and where $\sin(\beta_e)$ and $\cos(\beta_e)$ are given by Eq. (\ref{sincosbetae}) and $\Omega_e$ is given by Eq. (\ref{Omegae}).
Then, letting
\begin{eqnarray}
|\tilde{\psi}_{G,ne}(t)\rangle&=&e^{i(\beta_nI_y{\bm 1}_e+\beta_e{\bm 1}_n)J_y/\hbar}e^{i\omega t(I_z{\bm 1}_e+{\bm 1}_nJ_z)/\hbar}|\psi_{G,ne}(t)\rangle\nonumber\\
&=&e^{i(\beta_nI_y{\bm 1}_e+\beta_e{\bm 1}_nJ_y)/\hbar}\nonumber\\
& &\times e^{-it(\hat{\bm n}_n\cdot{\bm I}\Omega_n{\bm 1}_e+{\bm 1}_n\hat{\bm n}_e\cdot{\bm J}\Omega_e)/\hbar}\nonumber\\
& &\times|\psi_{G,ne}(0)\rangle,\nonumber\\
\end{eqnarray}
and then
\begin{eqnarray}
i\hbar\frac{\partial}{\partial t}|\tilde{\psi}_{G,ne}(t)\rangle&=&e^{i(\beta_nI_y{\bm 1}_e+\beta_e{\bm 1}_nJ_y)/\hbar}\nonumber\\
& &\times(\hat{\bm n}_n\cdot{\bm I}\Omega_n{\bm 1}_e+{\bm 1}_n\hat{\bm n}_e\cdot{\bm J}\Omega_e)\nonumber\\
& &\times e^{-it(\hat{\bm n}_n\cdot{\bm I}\Omega_n{\bm 1}_e+{\bm 1}_n\hat{\bm n}_e\cdot{\bm J}\Omega_e)/\hbar}|\psi_{G,ne}(0)\rangle\nonumber\\
&=&e^{i(\beta_nI_y+\beta_eJ_y)/\hbar}\nonumber\\
& &\times(\hat{\bm n}_n\cdot{\bm I}\Omega_n{\bm 1}_e+{\bm 1}_n\hat{\bm n}_e\cdot{\bm J}\Omega_e)\nonumber\\
& &\times e^{-i(\beta_nI_y{\bm 1}_e+{\bm 1}_n\beta_eJ_y)/\hbar}e^{i(\beta_nI_y{\bm 1}_e+{\bm 1}_n\beta_eJ_y)/\hbar}\nonumber\\
& &\times e^{-it(\hat{\bm n}_n\cdot{\bm I}\Omega_n{\bm 1}_e+{\bm 1}_n\hat{\bm n}_e\cdot{\bm J}\Omega_e)/\hbar}|\psi_{G,ne}(0)\rangle\nonumber\\
&=&(\Omega_nI_z{\bm 1}_e+\Omega_e{\bm 1}_nJ_z)|\tilde{\psi}_{G,ne}(t)\rangle,
\end{eqnarray}
after doing the rotations about the ${\bm y}$ axis for both the nuclear and electronic components, as described for the former in Sec. II.

\subsection{Time-dependent perturbations}

The exact wave function for the RWA to the problems of NMR and EPR   should be suitable for use as the starting point for future time-dependent perturbations, in order to more closely approximate the actual experimental situations.  A  time-dependent perturbation  of particular interest is $\hat{\delta{V}}_1(t)$ given by Eq. (\ref{deltaV1}).

For the simple case $a_p=0$ for $p\ge2$, this  would allow for a well-defined method to perturbatively treat the Hamiltonian $\hat{H}(t)=\omega_{n0}I_z+\omega_{n1}I_x\cos(\omega t)$ \cite{Layton} and its corresponding EPR equivalent, as in  early experiments \cite{RabiRamseySchwinger}. More general perturbations for a large class of periodic fields along $\hat{\bm x}$ can be treated
  for suitable $a_p$ and initial time $t_0$ \cite{Boas}.  This could treat square-wave and kinky (or triangular-wave) perturbations of various shapes, for example.

  This technique could also be useful for time-dependent perturbative treatments of current ``magic-angle'' spinning magnetic resonance experiments in biophysics and organic chemistry \cite{Hong}, for which the time-dependent perturbation $\hat{\delta{V}}_2(t)$ is given by Eq. (\ref{deltaV2}). 
  In such cases, the perturbation form can be obtained by initially starting with the strong field ${\bm H}_0=H_0[\hat{\bm x}\sin\theta_M+\hat{\bm z}\cos\theta_M]$, and the sample rotating at some small angle $\alpha$ about the ${\bm z}$ axis, so that the effective magnetic field from the sample rotation at the angular frequency $\omega$ is ${\bm H}_1(t)=H_1[\hat{\bm x}\cos(\omega t)+\hat{\bm y}\sin(\omega t)]$.  Then, by rotating ${\bm H}_0+{\bm H}_1(t)$ by $\theta_M$ about the $y$ axis, the effective Hamiltonian in the laboratory frame pictured in Fig. 1 of \cite{Hong}.
 Using the doubly-rotated bare wave function given by Eq. (\ref{doublyrotatedpsi}),  $\hat{H}_0$  is diagonal and independent of $t$, so that the time-dependent perturbations can be carried out in the interaction representation \cite{Gottfried,SN3,GY}.
 \begin{table}
\begin{tabular}{llclll}
isotope&$I$&$P$&$\mu_n/\mu_N$&$Q$ (b)&atomic \% \\
\hline
$^1$H&$1/2$&+&+2.79284734(3)&&99.985\\
$^2$H&1&+&+0.85743822(9)&+0.00285783(30)&0.0135\\
$^7$Li&$3/2$&-&+3.256427(2)&-0.0370(8) to&92.41\\
&&&+3.2564625(4)&-0.059(8)  (9 values)&\\
$^{13}$C&$1/2$&-&+0.7024118(14)&&1.1\\
$^{14}$N&1&+&+0.40376100(6)&+0.02044(3)&99.6\\
$^{15}$N&$1/2$&-&-0.28318884(5)&&0.38\\
$^{17}$O&$5/2$&+&-1.89379(9)&-0.02558(22)&0.037\\
$^{29}$Si&$1/2$&+&-0.55529(3)&&4.67\\
$^{31}$P&$1/2$&+&+1.13160(3)&&100\\
$^{33}$S&$3/2$&+&+0.6438212(14)&-0.0694(4)&0.75\\
$^{43}$Ca&$7/2$&-&-1.3173(6)&-0.408(8)&0.135\\
$^{57}$Fe&$1/2$&-&+0.0906&&2.119\\
$^{87}$Sr&$9/2$&+&-1.0936030(13)&+3.05(2)&7.00\\
$^{133}$Cs&$7/2$&+&+2.582025(3)&-0.00355(4)&100\\
&&&+2.5829128(15)&-0.00371(4)&\\
\hline
\end{tabular}
\caption{Table of the stable isotopes, spin quantum numbers $I$, parity $P$, nuclear magnetic moments $\mu_n$ relative to the nuclear magneton $\mu_N=(e\hbar)/(2m_p)=5.05783699(31)\times10^{-27}$ J/T, the electric quadrupole moment $Q$  in barns (1 b = 10$^{-28}$m$^2$), and atomic \% abundances, of the atoms in  adenine, cytosine, guanine, and thymine, three isotopes of possible chemical defect substitutions, Si, P and S, respectively for C, N and O, and three isotopes present in blood and bones.  $^7$Li and $^{133}$Cs are also listed.\cite{Stone,CIAAW,Pyykkoe} More recent and complete tables of the alkali are available \cite{Allegrini}.}
\end{table}

\section{Summary and conclusions}

Gottfried's elegant solution for the wave function of the rotating wave approximation to the nuclear magnetic resonance problem for general spin $I$ allows one to simply calculate the probability of a transition from a fully-occupied state to any other fully-occupied state \cite{Gottfried}.  But it does not allow one to calculate the probability of a transition from a fully general mixed state to another fully general mixed state. By diagonalizing Gottfried's wave function, we found a more useful exact wave function that allows such full generality of state transitions.  Since the time-dependent applied magnetic field in the rotating wave approximation also applies to the electrons in an atom or molecule, an equivalent version of our more useful nuclear wave function applicable to electron paramagnetic resonance for a general electronic total angular momentum operator $J$ was found.  With both more useful wave function components, we showed that a wave function form that is rotated about both the nuclear and electronic variables by the appropriate but greatly different nuclear and electronic rotation angles leads to a bare time-independent Hamiltonian $\hat{H}_0$ that is diagonal in both the nuclear and electronic variables.  This $\hat{H}_0$ allows for the calculation of the nuclear moment interactions proportional to sums of powers of ${\bm I}\cdot{\bm J}$ using standard time-independent perturbation theory. The elements of the matrix for the energies due to the magnetic dipole and electric quadrupole moment  in an RWA experiment  are found for general $I,J$.  Exact expressions for the  energies of the simplest case of $I=1$ and $J=\frac{1}{2}$, corresponding to the  $^2$P$_{\frac{1}{2}}$ and $^2$S$_{\frac{1}{2}}$ excited states of $^{14}$N are presented. In addition, this simple form of $\hat{H}_0$ allows for treatments of time-dependent perturbations in the standard interaction representation.

We propose  a new experimental technique to measure the nuclear moments of atoms and molecules.  One can construct the applied magnetic field to have a strong constant component in one direction and a weaker circularly-rotating component normal to it.  One would like to be able to vary the frequencies within at least an order of magnitude of both the nuclear magnetic resonance and the electron paramagnetic resonance frequencies, which differ by several orders of magnitude.  This may require two separate devices.

This work also applies to molecules with sufficiently high symmetry that the molecular total angular momentum in the absence of the applied field is quantized in terms of a set of  $J$ values. In molecules with high symmetry, such as benzene, this should be possible.  But in steric molecules such as a C atom bonded to four different atomic or molecular groups, such as H, Cl, NH$_2$, and OH, there is no center of symmetry, and hence no set of single $J$ values. The relevant point is that for molecules without high symmetry, such as each of the four amino acids in DNA, there is a mirror plane through the flat molecule, but none of these molecules exhibits rotational symmetry about the axis normal to that plane.  Thus, it is possible that mixed angular momentum states $J_1$ and $J_2$ could be present in them.  But how do we characterize the mixing parameter and its phase?  Those are questions for future
investigations.

 \section{Appendix}
  Gottfried used the notations $\hbar=1$, ${\bm J}$ for our ${\bm I}$, $\Omega$ for our $\omega_0$, $\lambda\Omega$ for our $\omega_1$, $\Delta$ for our $\Omega$,  $\varphi$ for our $\beta$, and his $\hat{H}_{RWA}(t)$ was opposite in sign to ours \cite{Gottfried}.   Although Gottfried was the first to correctly  diagonalize $\hat{H}_{RWA}(t)$, he wrote the wave function $|\psi(t)\rangle\equiv|t\rangle$ in our notation as
  \begin{eqnarray}
  |\psi(t)\rangle&=&e^{-iI_z\omega t/\hbar}e^{-i\hat{\bm n}\cdot{\bm I}\Omega t/\hbar}|\psi(0)\rangle,\\
  \hat{\bm n}&=&\hat{\bm z}\cos\beta+\hat{\bm x}\sin\beta,
  \end{eqnarray}
 where $\hat{\bm z}=\hat{\bm u}_z$ $\hat{\bm x}=\hat{\bm u}_x$.

 In order to clarify the equality of the matrix elements of Gottfried's wave function, Eq. (\ref{Gottfriedwavefunction}), and the present wave function, Eq. (\ref{psioftexact}), we shall demonstrate their equivalence for $I=1$.  In this case, it is easy to show that for $I=1$,  $\langle I,m'|({\bm I}\cdot{\hat{\bm n}}/\hbar)^3|I,m\rangle=\langle I,m'|{\bm I}\cdot{\hat{\bm n}}/\hbar|I,m\rangle$.  Letting $\zeta=\Omega t$ and $\hat{Q}=[I_z\cos(\beta)+I_x\sin(\beta)]/\hbar={\bm I}\cdot\hat{\bm n}/\hbar$, where $\hat{\bm n}=\hat{\bm z}\cos(\beta)+\hat{\bm x}\sin(\beta)$ and $\sin(\beta)$ and $\cos(\beta)$ are given in Eq. (\ref{sincosbeta}). Then letting $|I,m\rangle\equiv|m\rangle$ for $I=1$,
 \begin{eqnarray}
 d^{(1)}_{m',m}(\beta,\zeta)&\equiv&\langle m'|e^{-i\zeta\hat{Q}}|m\rangle\nonumber\\
 &=&\langle m'|{\bm 1}-i\sin(\zeta)\hat{Q}-[1-\cos(\zeta)]\hat{Q}^2|m\rangle\nonumber\\
 \end{eqnarray}
where the matrix elements satisfy
\begin{eqnarray}
d^{(1)}_{1,1}&=&
-i\sin(\zeta)\cos(\beta)+\frac{1}{2}\Bigl(\sin^2(\beta)+\cos(\zeta)[1+\cos^2(\beta)]\Bigr)\nonumber\\
&=&d^{(1)*}_{-1,-1},\\
d^{(1)}_{0,0}&=&\cos^2(\beta)+\sin^2(\beta)\cos(\zeta),\\
d^{(1)}_{1,0}&=&d^{(1)}_{0,1}=-\frac{\sin(2\beta)}{2\sqrt{2}}[1-\cos(\zeta)]\nonumber\\
& &-\frac{i}{\sqrt{2}}\sin(\beta)\sin(\zeta),\label{d10}\\
d^{(1)}_{1,-1}&=&d^{(1)}_{-1,1}=-\frac{1}{2}\sin^2(\beta)[1-\cos(\zeta)],\label{d1m1}\\
d^{(1)}_{0,-1}&=&d^{(1)}_{-1,0}=-d^{(1)*}_{1,0}.\label{d0m1}
\end{eqnarray}
It can also be shown that for general $I$,
\begin{eqnarray}
\langle m'|e^{-i\beta I_y/\hbar}e^{-i\zeta I_z/\hbar}e^{i\beta I_y/\hbar}|m\rangle&=&d^{(I)}_{m',m}(\beta,\zeta).
\end{eqnarray}
For $I=1$, each component of $d^{(1)}_{m',m}(\beta,\zeta)$ is identical to that calculated above. However, this expression can readily be evaluated for arbitrary $I$, and one does not require the individual matrix evaluations for each $I$ value, as outlined above for $I=1$.

 Gottfried did this for $I=\frac{1}{2}$ using the Pauli matrices \cite{Gottfried}. He showed
that the matrix elements could be written as
\begin{eqnarray}
D_{\hat{\bm n}}(\Omega t)&=&e^{-i(\hat{\bm n}\cdot{\bm\sigma})\Omega t/2}\nonumber\\
&=&{\bm 1}\cos(\Omega t/2)-i\hat{\bm n}\cdot{\bm\sigma}\sin(\Omega t/2)
\end{eqnarray}
in our notation. Since the standard rotation matrix involves time-independent rotations by the angle $\beta$ about the $\hat{\bm y}$ axis, for $I=\frac{1}{2}$, $d^{(1/2)}_{m',m}(\beta)$ has diagonal matrix elements $\cos(\beta/2)$ and off-diagonal matrix elements  $\mp\sin(\beta/2)$.  Then, he wrote that the probability of a transition from  state $m$ to  state $m'$, either for diagonal or off-diagonal states, would be $|d_{m',m}^{(1/2)}(\beta)|^2$  with $\beta$ given by $\sin(\beta/2)=\frac{\omega_1}{\Omega}\sin(\Omega t/2)$ \cite{Gottfried}.

 But he also wrote that this would apply for the general  $d^{(I)}_{m',m}(\beta)$ with $\beta$ also given  by $\sin(\beta/2)=\frac{\omega_1}{\Omega}\sin(\Omega t/2)$.  This statement may have confused some workers in NMR.  What he should have written was that the expression for the general diagonal or off-diagonal $(m',m)$ states for general $d^{(I)}_{m',m}(\beta)$ could be equated with the appropriate  elements of that time-dependent matrix.  Thus, for $I=1$, $d^{(1)}_{m',m}(\beta)$ has the off-diagonal elements $d^{(1)}_{1,0}=d^{(1)}_{0,-1}=-d^{(1)}_{0,1}=-d^{(1)}_{-1,0}=-\frac{1}{\sqrt{2}}\sin(\beta)$ and $d^{(1)}_{1,-1}=d^{(1)}_{-1,1}=\frac{1}{2}[1-\cos(\beta)]$ \cite{SN3}.  For each of these off-diagonal matrix elements, these $|d^{(1)}_{m',m}(\beta)|^2$ values are consistent with the $|d^{(1)}_{m',m}(\beta=\sin^{-1}(\omega_1/\Omega),\zeta=\Omega t)|^2$ presented in Eqs. (\ref{d10}) to (\ref{d0m1}).  We have shown analytically for the diagonal matrix elements $d_{0,0}^{(1)}=\cos(\beta)$ and $d_{1,1}^{(1)}=d_{-1,-1}^{(1)*}=\frac{1}{2}[1+\cos(\beta)]$ that this consistency applies to every matrix element.  We also found numerically that it is true for every $(m',m)$ combination, both diagonal and off-diagonal, of every spin $\frac{3}{2}$ combination  of   $|d^{(\frac{3}{2})}_{m',m}(\beta)|^2$.

 The transformed time-dependent perturbations are listed in the following.  We have for the nuclear cases,
\begin{eqnarray}
\widetilde{\hat{\delta V}}_{n1}(t)&=&-\omega_{n1}\sin(\omega t)\Bigl[I_y\cos(\omega t)\nonumber\\
& &+\sin(\omega t)\Bigl(I_x\cos(\beta_n)+I_z\sin(\beta_n)\Bigr)\Bigr]\nonumber\\
& &+\omega_{n1}\Bigl[\cos(\omega t)\Bigl(I_x\cos(\beta_n)+I_z\sin(\beta_n)\Bigr)\nonumber\\
& &-I_y\sin(\omega t)\Bigr]\sum_{p=2}^{\infty}a_p\cos\left[p\omega(t-t_0)\right],\\
\widetilde{\hat{\delta V}}_{n2}(t)&=&2\omega_{n1}\sin\theta\cos(\omega t)\nonumber\\
& &\times \Bigl[I_z\Bigl(\cos\theta\cos(\beta_n)-\sin\theta\sin(\beta_n)\cos(\omega t)\Bigr)\nonumber\\
& &-I_x\Bigl(\cos\theta\sin(\beta_n)+\sin\theta\cos(\beta_n)\cos(\omega t)\Bigr)\nonumber\\
& &-I_y\sin\theta\sin(\omega t)\Bigr].
\end{eqnarray}
For the electronic cases,
\begin{eqnarray}
\widetilde{\hat{\delta V}}_{e1}(t)&=&-\omega_{e1}\sin(\omega t)\Bigl[J_y\cos(\omega t)\nonumber\\
& &+\sin(\omega t)\Bigl(J_x\cos(\beta_e)+J_z\sin(\beta_e)\Bigr)\Bigr]\nonumber\\
& &+\omega_{e1}\Bigl[\cos(\omega t)\Bigl(J_x\cos(\beta_e)+J_z\sin(\beta_e)\Bigr)\nonumber\\
& &-J_y\sin(\omega t)\Bigr]\sum_{p=2}^{\infty}a_p\cos\left[p\omega(t-t_0)\right],\\
\widetilde{\hat{\delta V}}_{e2}(t)&=&2\omega_{e1}\sin\theta\cos(\omega t)\nonumber\\
& &\times \Bigl[J_z\Bigl(\cos\theta\cos(\beta_e)-\sin\theta\sin(\beta_e)\cos(\omega t)\Bigr)\nonumber\\
& &-J_x\Bigl(\cos\theta\sin(\beta_e)+\sin\theta\cos(\beta_e)\cos(\omega t)\Bigr)\nonumber\\
& &-J_y\sin\theta\sin(\omega t)\Bigr].
\end{eqnarray}

\section{Acknowledgments}

The authors acknowledge discussions with Luca Argenti, Bo Chen, James K. Harper, and Talat S. Rahman.
 R. A. K. was partially supported by the U. S. Air Force Office of Scientific Research (AFOSR) LRIR \#24RQCOR004, and the AFRL/SFFP Summer Faculty Program provided by AFRL/RQ at WPAFB.

 \end{document}